\documentclass[a4paper,fleqn,usenatbib,useAMS]{mnras}

\usepackage{newtxtext,newtxmath}
\usepackage{subfigure}
\usepackage[T1]{fontenc}
\usepackage{ae,aecompl}
\usepackage{multirow}
\usepackage{threeparttable}
\usepackage{longtable}

\usepackage{graphicx}	
\usepackage{amsmath}	
\usepackage{amssymb}	
\usepackage{pdflscape}	

\title[M-type stars recognized through hash learning]{Recognition of M-type stars in the unclassified spectra of LAMOST DR5 using a hash learning method}

\author[Y.-X. Guo et al.]{
Y.-X. Guo,$^{1,2}$
A.-L. Luo,$^{1,2, 3}$\thanks{E-mail: lal@nao.cas.cn}
S. Zhang,$^{1,2}$
B. Du,$^{1}$
Y.-F. Wang,$^{1}$
J.-J. Chen,$^{1}$
\newauthor
F. Zuo,$^{1,2}$
X. Kong,$^{1}$
Y.-H. Hou$^{4}$
\\
$^{1}$Key Laboratory of Optical Astronomy, National Astronomical Observatories, Chinese Academy of Sciences, Beijing 100012, China\\
$^{2}$University of Chinese Academy of Sciences, Beijing 100049, China\\
$^{3}$Department of Physics and Astronomy, University of Delaware, Newark, DE 19716, USA\\
$^{4}$Nanjing Institute of Astronomical Optics \& Technology, National Astronomical Observatories, Chinese Academy of Sciences,\\
Nanjing 210042, China
}

\date{Accepted 2019 February 05. Received 2019 February 03; in original form 2018 July 08}
\pubyear{2018}

\begin{document}
\label{firstpage}
\pagerange{\pageref{firstpage}--\pageref{lastpage}}
\maketitle

\begin{abstract}
Our study aims to recognize M-type stars which are classified as ``UNKNOWN'' due to bad quality in Large sky Area Multi-Object fibre Spectroscopic Telescope (LAMOST) DR5 V1. A binary nonlinear hashing algorithm based on Multi-Layer Pseudo Inverse Learning (ML-PIL) is proposed to effectively learn spectral features for the M-type star detection, which can overcome the bad fitting problem of template matching, particularly for low S/N spectra. The key steps and the performance of the search scheme are presented. A positive dataset is obtained by clustering the existing M-type spectra to train the ML-PIL networks. By employing this new method, we find 11,410 M-type spectra out of 642,178 ``UNKNOWN'' spectra, and provide a supplemental catalogue. Both the supplemental objects and released M-type stars in DR5 V1 are composed a whole M type sample, which will be released in the official DR5 to the public in June 2019, All the M-type stars in the dataset are classified to giants and dwarfs by two suggested separators: 1) color diagram of $H$ versus $J-K$ from 2MASS; 2) line indices CaOH versus CaH1, and the separation is validated with HRD derived from $Gaia$ DR2. The magnetic activities and kinematics of M dwarfs are also provided with the EW of H$\alpha$ emission line and the astrometric data from $Gaia$ DR2 respectively.
\end{abstract}




\section{Introduction}
\label{sec:info}
M-type stars are becoming dominant targets for research on the structural evolution and kinematics of the local Milky Way. M giants and M dwarfs have similar spectral features, both with strong molecular characteristics. M giants are red-giant-branch (RGB) stars with low surface temperature and high luminosity in the late-phase of stellar evolution. Their luminous nature allows us to use these stars as good tracers to study the outer Galactic halo and distant substructures \citep{2015RAA....15.1154Z}. M dwarfs, main-sequence stars with $M_*\sim0.075$--$0.6M$\sun \citep{2000ApJ...542..464C}, are the dominant stellar constituent in the solar neighborhood and probably the Galaxy \citep{1994AJ....108.1437H,2006AJ....132.2360H,2010AJ....139.2679B,1995ApJ...441...51S,2003PASP..115..763C}. They are very useful sources for studying and probing the lower end of  the Hertzsprung-Russell diagram (HRD), even down to the hydrogen-burning limit. More and more M dwarf samples enable us to deep the understanding of their fundamental properties just like what we knew about the massive stars. For example, some of the key astrophysical interrelated topics have been exploring, including the precise relationship between mass and radius \citep{2014ApJ...789...53F,2014MNRAS.441.2111J,2017AJ....154..100H}, the mass-luminosity relation \citep{1993AJ....106..773H,2000A&A...364..217D,2010A&ARv..18...67T,2016AJ....152..141B}, rotation and angular momentum \citep{2011ASPC..448..505S,2017ApJ...837...96H}, magnetic activity \citep{2012LRSP....9....1R,2014ApJ...789...53F,2017ApJ...849...36Y}, complex atmospheric parameters and dust settling in their atmospheres, and age dispersion within populations \citep{2017ApJ...851...26V,2017MNRAS.465..760B}.  

The largest spectroscopic data bases of M-type stars were from multi-object spectroscopic surveys such as the Sloan Extension for Galactic Understanding and Evolution (SEGUE) \citep{2009AJ....137.4377Y} and the LAMOST Experiment for Galactic Understanding and Exploration (LEGUE) \citep{2012ASPC..458..405N}. Besides the formal data releases of the surveys, specific M dwarf catalogs were also presented by astronomers. An M dwarf catalogue of SDSS including more than 70,000 stars \citep{2011ASPC..448.1407W}, and two M dwarf catalogues of LAMOST were published \citep{2014AJ....147...33Y,2015RAA....15.1182G}. Considering the intrinsic low brightness of M dwarfs and the large distance of M giants, however, many low S/N M-type spectra has not  been recognized in these surveys.

LAMOST DR5 V1 have released more than 9 million spectra including 640,000 ``UNKNOWN''  spectra (not classified by the LAMOST pipeline \citep{2015RAA....15.1095L}).  Some of these ``UNKNOWN'' spectra, mostly with low S/N, are valuable for astronomical research. For example, \citet{2017RAA....17...32H} identified eight quasars from the LAMOST DR3 ``UNKNOWN'' spectra in the area of the Galactic anti-center of $150^\circ\leq l \leq210^\circ$ and $|b|\leq 30^\circ$. By applying a machine learning method, \citet{2018ApJS..234...31L} recognized a total of 149 carbon stars that were misclassified as ``UNKNOWN'' in LAMOST DR4. \citet{2018MNRAS.477.4641R} published a  catalog of White Dwarf Main Sequence binaries based on DR5 V1 dataset, several of which were classified as ``UNKNOWN'' by the LAMOST pipeline. 

The classification method of LAMOST pipeline is based on template matching, in which each observed spectrum is cross-matched with a set of templates to calculate chi-square values. The template which corresponds the smallest value suggests the class that the object belongs to. Sometimes, the chi-square value of the best-fitted template for a low S/N spectrum has too low confidence, which makes the pipeline refuse to judge and labels its class as ``Unknown".  Other than template matching, the Query based machine learning methods are specifically `similarity search'  algorithm which can retrieve objects in a database with a specific pattern ignoring irrelevant noise. The Approximate Nearest Neighbor Search (ANNS)  is a commonly used Query method, and the hash learning technique is one of the most widely used ANNS algorithm \citep{Wang2016A}.  The basic idea of the hashing-based search techniques is to learn the relationships which map the high-dimensional raw data to the compact binary codes (series of digits consisting of 0 and 1), and then to retrieve the nearest neighbors of the query pattern using the Hamming distance (frequently used for representing the distance between two binary code) in the binary code space. Consequently, searching in the hash code space is extremely efficient both in time and memory consuming. 

\begin{figure}
    \includegraphics[width=\columnwidth]{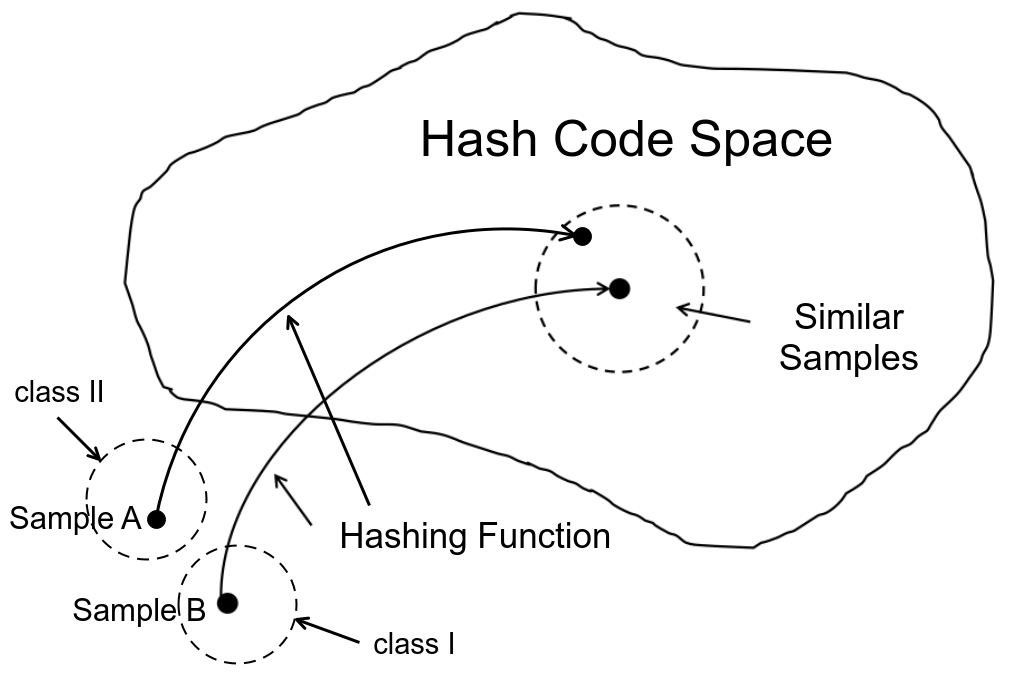}
    \caption{Principle schematic diagram of the hashing learning search. Members of two different classes in original space might be similar in the hashing code space through a hash mapping which can be a coding network obtained via deep learning.}
    \label{fig:fig1}
\end{figure}

A schematic diagram of a hash learning search is shown in Fig.\ref{fig:fig1}. Many hash methods, including Locality Sensitive Hashing (LSH) \citep{Andoni2006Near}, Spectral Hashing (specH) \citep{Weiss2008Spectral}, Iterative Quantization (ITQ) \citep{Gong2011Iterative}, Spherical Hashing (SpH) \citep{Heo2012Spherical} etc., have been intensively studied and widely used in many different fields, and their advantages and weaknesses have also been deeply investigated \citep{Bondugula_surveyof, Wang2016Learning}. In this paper, we employ Semantic Hashing (SH) \citep{Salakhutdinov2009Semantic} to construct a deep hash learning model to search for M-type spectral pattern through learning  hidden binary features and reconstructing the input data.  However, to train such a deep generative model often requires multiple iterations, which suggests that it is not only extremely time-consuming while dealing with large amount of data but also needs to set parameters repeatedly depends on experience rather than theoretical basis. 

The appearance of pseudo inverse learning (PIL) \citep{Guo2017Autoencoder} shed a new light on the deep learning technique because PIL is actually an supervised learning algorithm for training a single hidden layer feedforward neural network which do not need to tune the hidden layer parameters once the number of hidden layer nodes is determined. The weight and bias vectors between the input layer and the hidden layer are randomly generated, and these are independent of the training samples and the specific applications \citep{Pal2015A}. In this study we build a multilayer PIL (ML-PIL) to fulfill the hash learning process, so as to search M-type stars in the ``UNKNOWN'' spectra of LAMOST DR5 V1. 

There are methods to separate giants from dwarfs for M-type stars based on colors, spectral indices and proper motions etc. \cite{1988PASP..100.1134B} proposed a color discrimination method. In their study, M giants and M dwarfs are distributed around different loci in the [$J-H$, $H-K$] color-color diagram, which are mainly caused by the differences in the opacity of molecular bands of H$_{2}$O \citep{1998A&A...337..321B}. Because M giants have relatively larger distances and smaller proper motions, a reduced proper motion method was used to separate M giants from M dwarfs \citep{2011AJ....142..138L}. By comparing the spectra of M dwarfs and M giants, several gravity-sensitive molecular and atomic spectral indices were selected to determine the luminosity class \citep{2012ApJ...753...90M}. Recently, a new photometric method combining 2MASS and WISE photometry was used to recognize M giant spectra in LAMOST dataset \citep{2015RAA....15.1154Z}. The strength ratio of TiO band to CaH band varies with surface gravity. \citet{1995AJ....110.1838R} defined the TiO5, CaH2, and CaH3 spectral indices,  and \citet{2015RAA....15.1154Z} used the aforementioned indices to distinguish M giants from M dwarfs. In addition, other methods, such as Mg2 versus $g-r$ was used for the separation \citep{2008AJ....136.1778C}. We compare different giant/dwarf separation schemes and suggest two additional separation indicators with more correctness.  

The paper is organized as follows: section \ref{sec:data} briefly introduce the spectral data used in the paper along with the spectra preprocessing; Section \ref{sec:SCHEME} presents the ML-PIL-based hash learning scheme, the construction of positive and negative samples, the model training and the performance evaluation of the method on real spectral data,  and then the application of ML-PIL in searching for M-type stars in LAMOST DR5 V1 ``UNKNOWN'' dataset; Section \ref{sec:discussion} compares different giant/dwarf separation schemes for M-type stars and suggests two useful indicators following by investigation of the activity and kinematics of the whole M dwarf sample in DR5; The final section summarizes the work of this paper and envisions potential future work.

\section{data and preprocessing}
\label{sec:data} 

LAMOST is a 4-m reflecting Schmidt telescope with a large field of view (FoV) of 5 degrees in diameter. It has 4,000 fibers mounted on its focal plane and 16 spectrographs with 32 CCD cameras, so that it can simultaneously observe up to 4,000 objects \citep{2012RAA....12.1197C}. The raw CCD data are reduced and analyzed by the LAMOST data pipelines, which consists of a 2D pipeline and a 1D pipeline. The primary functions of the 2D pipeline include bias calibration, flat field correction, spectral extraction, sky subtraction, wavelength calibration, flux calibration, sub-exposures combination, etc. The calibrated spectra from the 2D pipeline are then fed to the 1D pipeline which performs spectral classification and parameter determination based on template matching and chi-square criteria \citep{2015RAA....15.1095L}.

Until July 2017, LAMOST has completed its five-year regular survey. The LAMOST DR5 V1 includes 9,017,844 spectra of stars, galaxies, quasars, and unrecognized objects. These spectra cover the wavelength range from 3690 to 9100\AA\ with a resolution of  $R \sim 1800$ at the wavelength of 5500\AA.

\subsection{``UNKNOWN'' data from LAMOST DR5}

\begin{figure}
	\subfigure{\includegraphics[width=\columnwidth]{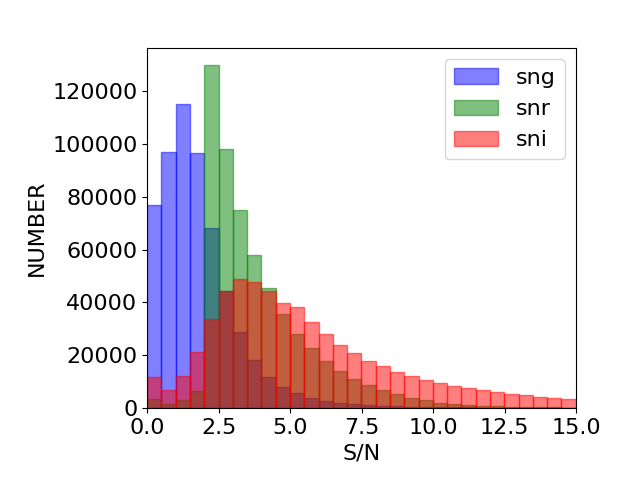}}
	\subfigure{\includegraphics[width=\columnwidth]{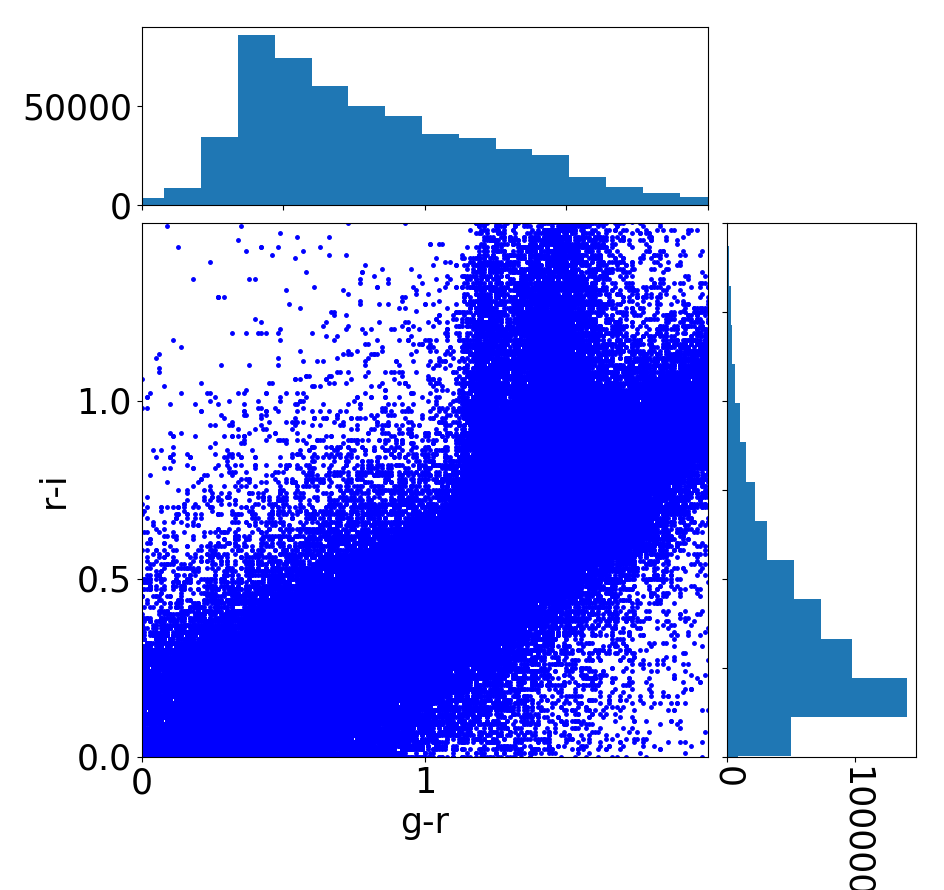}}
    \caption{S/N (top panel) and color g-r vs r-i (bottom panel) distribution of LAMOST ``UNKNOWN'' data. Most of late type M dwarfs are expected to be existed, which locate in the upper right region of the color-color diagram.}
    \label{fig:fig2}
\end{figure}

Among the 9 million spectra in LAMOST DR5 V1, 642,178 unrecognized spectra were labeled as ``UNKNOWN''. During the classification process of 1D pipeline, a spectrum is classified as ``UNKNOWN'' if the confidence of the classification result is lower than a given threshold value, e.g., the chi-square value of the best-match result is greater than a certain value, or the target spectrum has almost equal similarities to multiple dissimilar templates. These problems occur in multi-template matching process, which we will refer to as the multi-template matching problems, mostly owing to the low spectral S/N (see top panel of Fig. \ref{fig:fig2}).
 
The lower panel of Fig. \ref{fig:fig2} gives part of ``UNKNOWN'' objects in color-color diagram. The M-type star candidates should be located in the upper right region of this panel. Due to the intrinsic low luminosity of late-type M dwarfs, most of them have low S/N spectra, which are expected to be classified as ``UNKNOWN'' objects. To efficiently recognize M-type spectra from the 642,178 ``UNKNOWN'' spectra by using a more noise-insensitive approach than the multi-template matching problem of 1D pipeline, we choose an approximate proximity search method based on deep learning model, which can combine the low-level features layer by layer to obtain more abstract high-level feature expression, and then discover the inherent and essential feature representation of complex data.

\subsection{Data preprocessing}
\label{sec:preprocess}
LAMOST spectra cover the wavelength range from 3690 to 9100\AA\ with a resolution $R \sim 1800$ at the wavelength of 5500\AA. First, each spectrum is rebinned onto the same-wavelength space with a fixed step length. Then, we normalize the spectra by re-scaling the fluxes to eliminate scale differences among the raw spectra. For a given spectral set denoted by $S=\{x^{i}\}_{i=1}^{N}$, the vector $x^{i}=(x_1^{i},x_2^{i},\dots,x_p^{i})^{T}\in R^{p}$ represents a spectrum, in which $x$ is the flux at a given wavelength. The normalization is performed as

\begin{equation}
    \tilde{x_{n}^{i}} = \frac{({\rm MAX}-{\rm MIN})(x_{n}^{i}-\min{\{x^{i}\}})}{(\max{\{x^{i}\}} - \min{\{x^{i}\}})} + {\rm MIN},
	\label{eq:t}
\end{equation}

where MAX and MIN indicate the maximum and minimum values after the normalization, use 1 and 0 for simplicity. The $\max\{\cdot\}$ and $\min\{\cdot\}$ return the maximum and minimum element in a given vector, respectively. For each spectrum, we obtain the normalized one denoted as $\tilde{x}^{i}=(\tilde{x}_1^{i},\tilde{x}_2^{i},\dots,\tilde{x}_p^{i})^{T}$.

\section{ML-PIL based hashing scheme and application}
\label{sec:SCHEME} 
ML-PIL based hashing scheme can be divided into two stages: the deep hashing learning model training stage and the ANNS query stage.

In the model training phase, we construct a deep hash learning model to project all the target data into a feature space, then we encode the final feature representations of the last hidden layer's outputs into ``fingerprints''. For a well-trained ML-PIL-based hash network, we can get the corresponding ``fingerprints'' using the query sample of Section \ref{sec:positive} as input data. Similarly, we can get the ``fingerprints'' of the ``UNKNOWN'' spectra in DR5. We organize the description of this model construction into several subsections including framework of ML-PIL, hashing encoding scheme, positive sample through clustering, negative sample selection, model training and performance evaluation etc., from Section \ref{framework} to \ref{training}.

In the second ANNS query stage, for any given ``query'', we search for the similar spectra from the ``UNKNOWN'' data by calculating their similarities. The similarity between the query sample and each ``UNKNOWN'' spectrum is calculated by measuring the Hamming distance in the feature code space. The less distance of a sample to a coded query spectrum in the hash space suggests it is more similar to that query spectrum. The Section \ref{sec:application} illustrates the aforementioned query stage.  

\subsection{ML-PIL framework}
\label{framework}
\begin{figure}
	\centering
	\includegraphics[width=1.05\columnwidth]{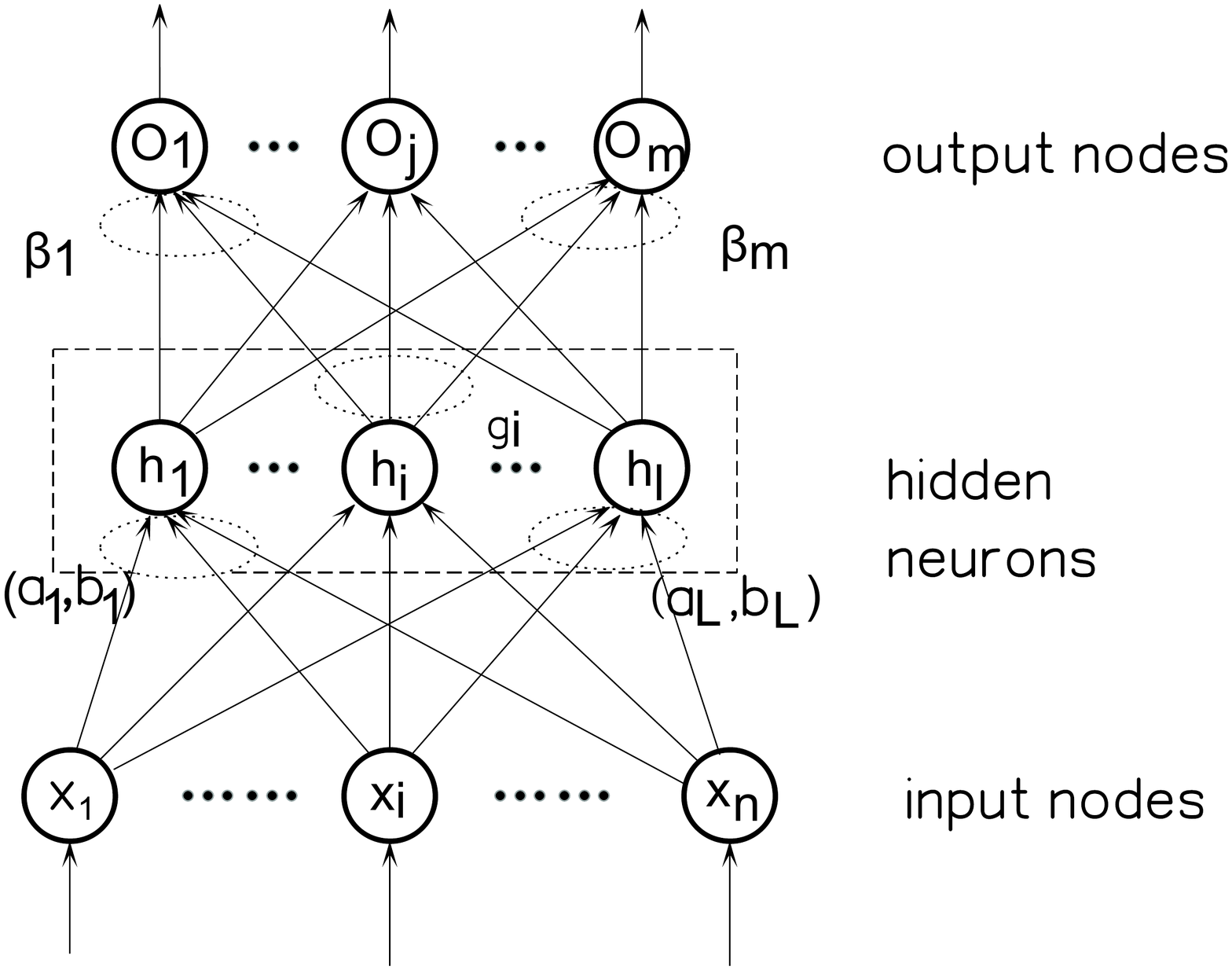}
    \caption{ The structure of single hidden layer feed-forward neural network (SLFN) .}
    \label{fig:fig3}
\end{figure}

ML-PIL is a hierarchical network structure based on pseudo inverse learning, and it is stacked with several single hidden layer neural networks. For a given single hidden layer neural network in Fig. \ref{fig:fig3}, we can get a single layer auto-encoder by training such that the output is approximately equal to input.  By stacking several aforementioned single layer autoencoders, we can get the multilayer autoencoder. Once a multilayer autoencoder is trained, the binary hash code of any sample is obtained from the deepest hidden layer. However, these complex models require iterative parameter adjustments and hence are computationally expensive.

To overcome the computational complexity of multilayer autoencoder, a PIL algorithm is introduced exploiting the advantages of its random orthogonal feature mapping to speed up learning. PIL is actually a supervised learning algorithm for training a single hidden layer feed-forward neural network(SLFN). The basic idea is to find a set of orthogonal vector bases using the nonlinear activation function to make the output vectors of the hidden layer neurons orthogonal. Then the weights of the output connection of the network are approximately solved by calculating the pseudo inverse. The PIL algorithm uses only basic matrix operations to calculate the analytical solution of the optimization objective \citep{Wang2016A,Wang2017Deep}.  It  does not need iterative optimization and parameter adjustment. Therefore,  its efficiency is much higher than that of the gradient descent based algorithms. Here, we give a detailed introduction for the PIL algorithm.

In Fig. \ref{fig:fig3}, suppose that $\{(X_{i},t_{i})\}(i=1,\dots,N)$ denotes the sample set, where $X_{i}=[x_{i1},x_{i2},\dots,x_{in}]^{T} \in R^{n}$ is a spectrum with n dimensions and $t_{i}=[t_{i1},t_{i2},\dots,t_{im}]^{T} \in R^{m}$ is the  target  label corresponding to $X_i$. The input $X_i$ is mapped to L-dimensional PIL random feature space, and the network output $o_{j}$ is 

\begin{equation}
    \sum_{i=1}^{L}{\beta_{i}g_{i}(W_{i}\centerdot X_j+b_i)}, j=1,\cdots,N
	\label{eq:oj}
\end{equation}
where $g_{i}$ is the activation function, $W_i=[a_{i1},\dots,a_{iL}]^{T}$ is the input weight vector, $\beta=[\beta_{1},\dots,\beta_{L}]^{T}$ is the output weight matrix between the hidden node and the output node, $b_i$ is the bias of the input matrix.
The aim is to find the optimal weight matrix to minimize the loss function 
\begin{equation}
    \sum_{j=1}^{N}||o_j-t_j||=0
	\label{eq:E}
\end{equation}

This problem can be expressed as
\begin{equation}
    H\beta = T
	\label{eq:t}
\end{equation}
where $H$ is the hidden layer node output. This nonlinear mapping $H$ is defined by
\begin{equation}
\begin{split}
    &H(W_1,\dots,W_L,b_1,\dots,b_L,X_1,\dots,X_L)\\
    =&\left[\begin{array}{ccc}
    g_1(W_1\centerdot X_1+b_1) & \cdots & g_1(W_L\centerdot X_1+b_L)\\
    \vdots & \dots & \vdots\\
    g_L(W_1\centerdot X_N+b_1) & \cdots & g_L(W_L\centerdot X_N+b_L)
    \end{array} 
    \right ]_{N\times L}
    \end{split}
	\label{eq:weight}
\end{equation}

The objective of optimization can be converted to minimize the loss function
\begin{equation}
    E=\sum_{j=1}^{N}(\sum_{i=1}^{L}\beta_{i}g_i(W_{i} \cdot X_{j}+b_{i})-t_{j})^{2}
	\label{eq:E}
\end{equation}

In the PIL algorithm, once the bias and the input weight of hidden layer is determined, the output matrix of hidden layer is uniquely determined. The training of the single hidden layer neural network can be transformed into solving a linear system. We can get the output weight $\beta$ from
\begin{equation}
    \hat{\beta} = H^{\dagger} T,
	\label{eq:hb}
\end{equation}
where $H$ is the Moore-Penrose generalized inverse of matrix $H$.

\begin{figure*}
    \includegraphics[width=\textwidth]{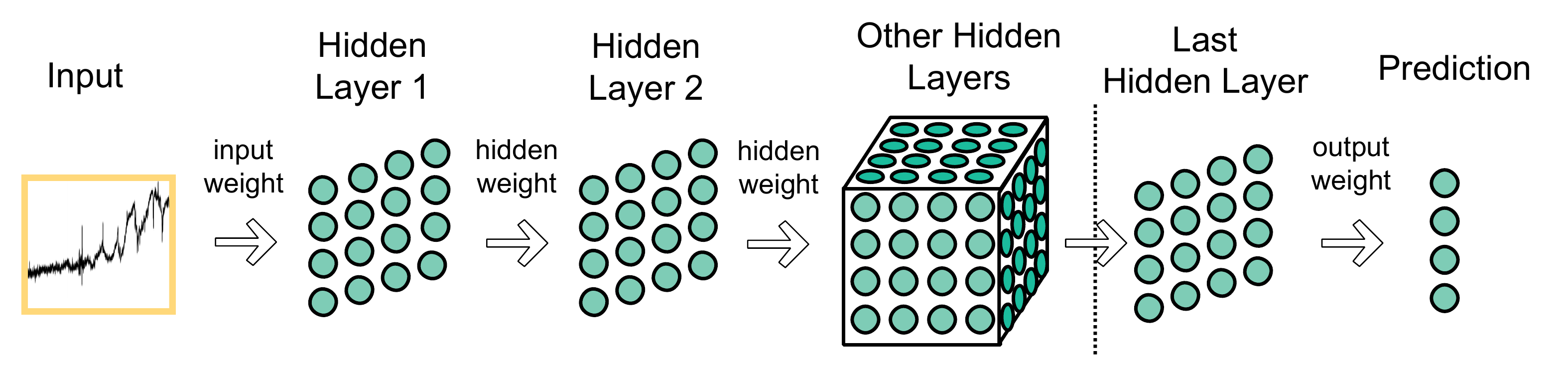}
    \caption{Framework of ML-PIL. Each hidden layer is trained separately. The last layer of ML-PIL can conduct a baseline PIL classification or regression.}
    \label{fig:fig4}
\end{figure*}

PIL is modified as follows to get PIL AutoEncoder (PIL AE) so as to perform unsupervised feature representation: input data are used as output data $T=X$. ML-PIL is derived from multiple stacks of PIL AEs. Each PIL AE is trained separately. The output of the hidden layer of the previous PIL-AE is connected to the input of the latter PIL-AE. The layer by layer trained PIL-AEs are then stacked into a ML-PIL (see Fig. \ref{fig:fig4}). The output of the last hidden layer is used to do hash mapping. 

\subsection{Proposed hashing scheme}

As described in the previous subsection, the feature expression can be learned from the last hidden layer of ML-PIL. These features can be projected into the hash code space through hash mapping to obtain the ``fingerprint''. The ``fingerprint'' is a binary number consisting of a series of 0 or 1. A perfect hash mapping should have the following properties simultaneously: (1) Similar samples should be mapped to similar hash codes (usually called similarity-preserving or coding consistency). (2) The hash codes should be ``balanced'' (usually called coding balance), which means that, for each bit in the code, half of the samples are mapped to 1 and the other half are mapped to 0 (or $-1$). (3) All bits should be independent of each other.  

\begin{figure}
	\centering
	\includegraphics[width=0.9\columnwidth]{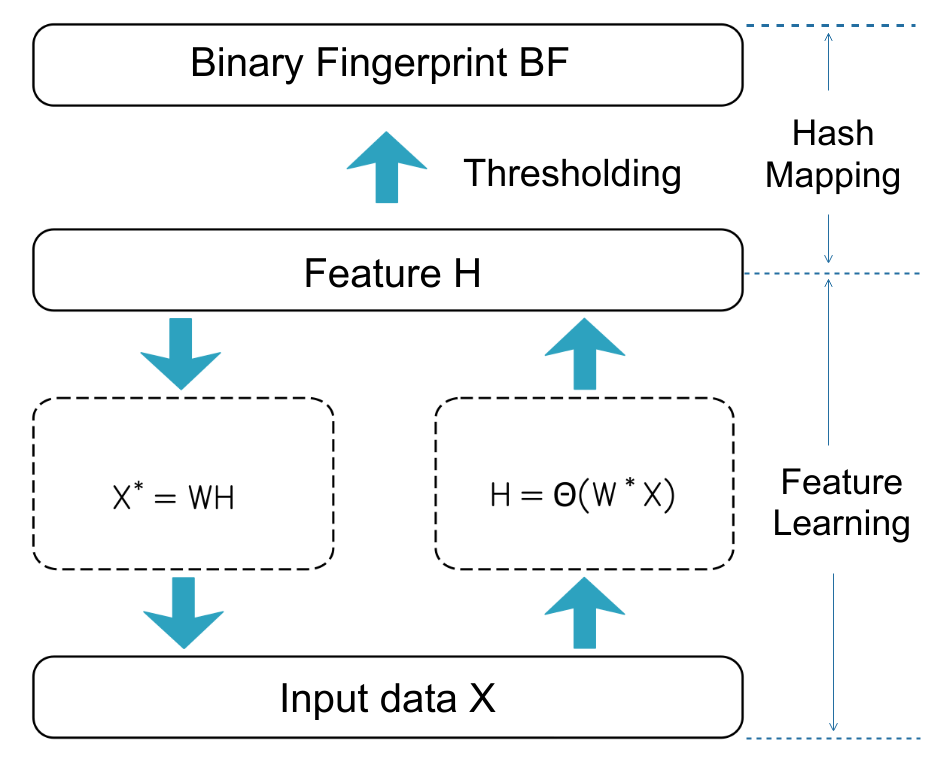}
    \caption{The procedure of learning and hash coding for features for a single hidden layer PIL-AE. $W^{\ast}$ denotes the random weight matrix; $X^{\ast}$ is the reconstruction of the input data. H is the learned feature through PIL-AE. }
    \label{fig:fig5}
\end{figure}

Fig. \ref{fig:fig5} illustrates the procedure of learning and hash coding for features.  We define a threshold with which the features H are made binary. To be specific, we choose the median value of each feature dimension as the threshold. Then the feature values that are greater than the threshold are mapped to 1;  otherwise, they are mapped to 0. By doing so, the learned binary codes are guaranteed to be ``balanced''. 

\subsection{Positive samples}
\label{sec:positive}

The size of training set for any Machine Learning (ML) algorithm depends on the complexity of the algorithm, while for PIL-ML based hashing scheme, thousands of positive samples are demanded to represent M-type spectra which embrace all kinds of subtypes, luminosity classes and various S/Ns especially low S/N ones since the ``UNKNOWN'' data have universally low S/N as shown in the top panel of Fig. \ref{fig:fig2}. Therefore, we cluster the released M-type stars in LAMOST DR5 V1 to select various positive samples from each cluster.

Before clustering, all the M-type spectra are shifted to rest frames, then two machine learning methods are adopted which are Balanced Iterative Reducing and Clustering Using Hierarchies (BIRCH) \citep{Zhang1999An} and K-means \citep{Arthur2007k}. The BIRCH algorithm builds a tree called the Characteristic Feature Tree (CFT) for the given data. It incrementally clusters the data points, uses a fraction of the dataset memory, and updates the clustering decisions when new data comes in. The K-Means algorithm clusters data by trying to separate samples in n groups of equal variance, minimizing a criterion known as the inertia or within-cluster sum-of-squares. It has been widely used in many different fields \citep{0004-637X-763-1-50,2014AJ....147..101W}.

\begin{figure}
	\subfigure{\includegraphics[width=\columnwidth]{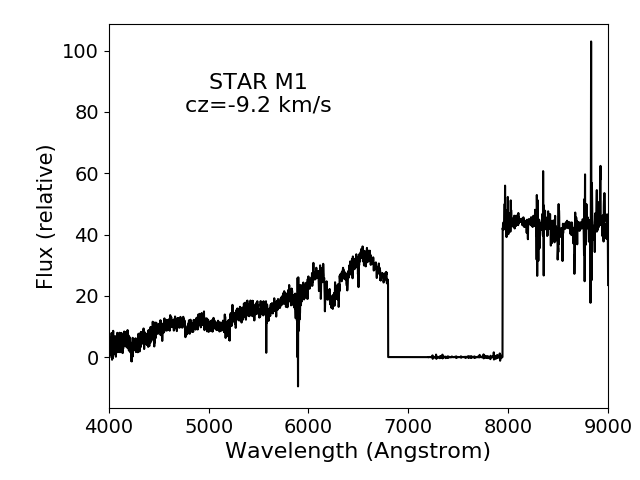}}
	\subfigure{\includegraphics[width=\columnwidth]{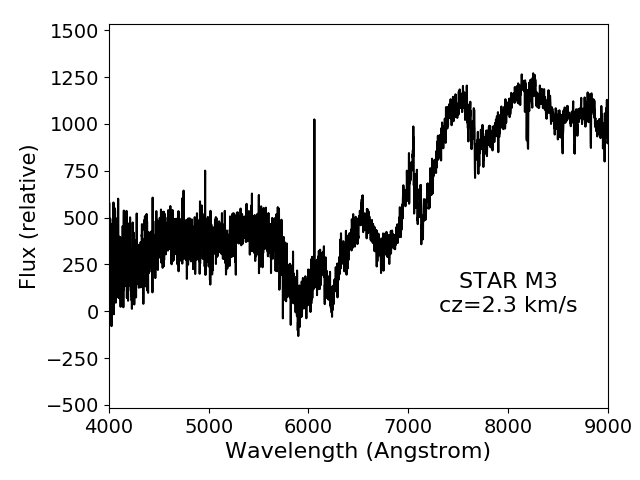}}
    \caption{Examples of defective spectra eliminated from positive samples.}
    \label{fig:fig6}
\end{figure}

First, the BIRCH algorithm is adopted to cluster the 529,629 M dwarf spectra in LAMOST DR5 V1 into 50 groups. The Principal Component Analysis \citep{Jolliffe2002Principal} is applied to reduce the dimensions of the spectra. Second, for each group, 20 sub-clusters are obtained using K-Means. While for M giants, 80 clusters are obtained only using K-Means algorithm. Thus, we initially obtain 1,080 average spectra for all cluster centers. After manually inspection, 23 defective spectra with flux gaps (see Fig.~\ref{fig:fig6}) are abandoned, and then 6,699 spectra are randomly selected in reserved 1057 clusters.  Supplementing 38 template spectra used in the LAMOST pipeline, 6,737 M-type positive spectra are ultimately assembled. 

\begin{figure}
	\centering
	\includegraphics[width=\columnwidth]{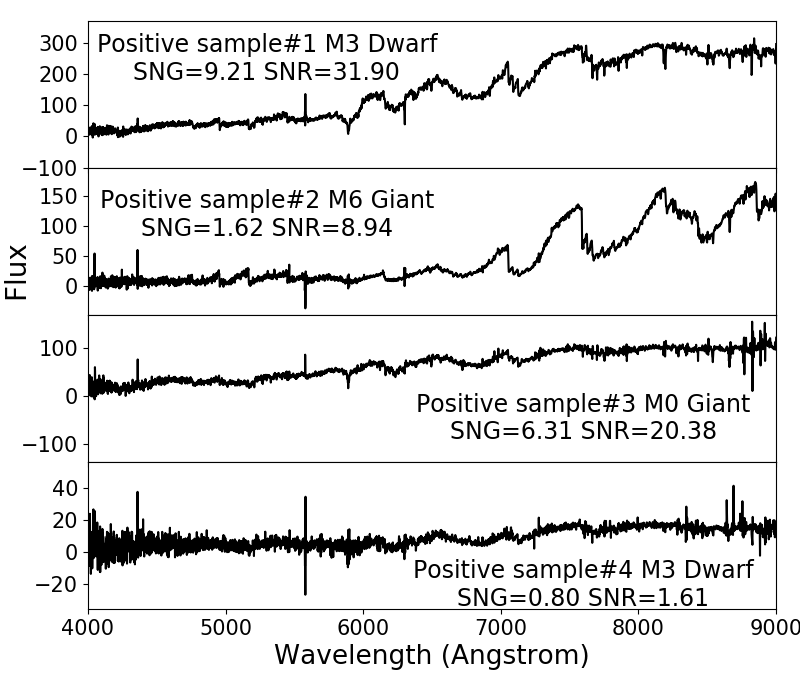}
    \caption{Example spectra in the positive samples with different subtypes, luminosity classes and S/Ns.}
    \label{fig:fig7}
\end{figure}

As shown in Table ~\ref{tab:table1}, the positive samples of the 6,737 M-type spectra include 10 M-type subclasses, luminosity classes (dwarf or giant), and wide range of S/Ns.  We present four typical positive samples shown in Fig.~\ref{fig:fig7}, three high S/N spectra and one low S/N  spectrum including an early giant, a late giant, and two early dwarfs.
 
\begin{table}
	\centering
	\caption{Main information of the finally combined query library.}
	\label{tab:table1}
	\begin{threeparttable}
	\begin{tabular}{cccccc} 
		\hline
		Spectral & Luminosity & \multicolumn{4}{c}{Numbers of different S/N\tnote{b}}\\		
        type & class\tnote{a} & $<$5 & 5$\sim$10 & 10$\sim$30 & $>$30\\ \hline
		\multirow{2}{*}{M0} & g & 107 & 59 & 61 & 337\\
		& d & 329 & 460 & 732 & 987 \\
		\hline
		\multirow{2}{*}{M1} & g & 37 & 10 & 16 & 104\\
		& d & 132 & 171 & 235 & 269 \\
		\hline
		\multirow{2}{*}{M2} & g & 69 & 12 & 23 & 83\\
		& d & 321 & 202 & 259 & 198 \\
		\hline
		\multirow{2}{*}{M3} & g & 87 & 8 & 6 & 10\\
		& d & 517 & 146 & 146 & 113 \\
		\hline
		\multirow{2}{*}{M4} & g & 28 & 10 & 29 & 73\\
		& d & 76 & 9 & 0 & 0 \\
		\hline
		\multirow{2}{*}{M5} & g & 5 & 7 & 14 & 31\\
		& d & 6 & 3 & 0 & 0 \\
		\hline
		\multirow{2}{*}{M6} & g & 14 & 13 & 17 & 47\\
		& d & 4 & 2 & 4 & 4 \\
		\hline
		\multirow{2}{*}{M7} & g & 8 & 4 & 5 & 34\\
		& d & 1 & 0 & 0 & 0 \\
		\hline
		\multirow{2}{*}{M8} & g & 11 & 1 & 3 & 4\\
		& d & 3 & 0 & 0 & 0 \\
		\hline
		\multirow{2}{*}{M9} & g & 13 & 0 & 1 & 0\\
		& d & 4 & 3 & 0 & 0 \\
		\hline
	\end{tabular}
	\begin{tablenotes}
	\item[a] g denotes giant; d denotes dwarf. 
	\item[b] S/N here refers to the r-band S/N value. 
	\end{tablenotes}
	\end{threeparttable}
\end{table}

\subsection{Negative samples}
\label{sec:negative}  
The 10,000 negative samples are randomly selected and visually confirmed as non-M-type from ``UNKNOWN'' spectra in LAMOST DR5 V1. Another 5000 non-M-type spectra are randomly selected from the data release  with known class and shifted to rest frames. Then, we totally obtain 15,000 negative samples.

\subsection{Model training and performance evaluation}
\label{training}
The aforementioned total 21, 737 positive and negative samples are used to train and validate the designed ML-PIL model. The  ML-PIL model comprises of three hidden layers, since \cite{Wang2017Deep} demonstrated that more hidden layers do not help much for improving performance. The Sigmoid activation function is selected for each hidden layer. The length of hash code (``fingerprint'') which is derived from the feature learned through ML-PIL would affect the performance of the ANNS, so that an appropriate code length should be decided via the performance evaluation. 

We use ``Accuracy'' and ``Recall'' to evaluate the performance of ML-PIL hashing searching. The ``Accuracy'' is defined as
\begin{equation}
    \rm Accuracy = \frac{TP}{TP + FP}
	\label{eq:ACCURACY}
\end{equation}
where TP denotes the number of the true positive samples in the result of query. While the FP is the number of the false positive samples.

The ``Recall'' is defined as
\begin{equation}
    \rm Recall = \frac{TP}{TP + FN}
	\label{eq:RECALL}
\end{equation}
where FN denotes the number of the false negative samples. 

\begin{figure}
    \includegraphics[width=\columnwidth]{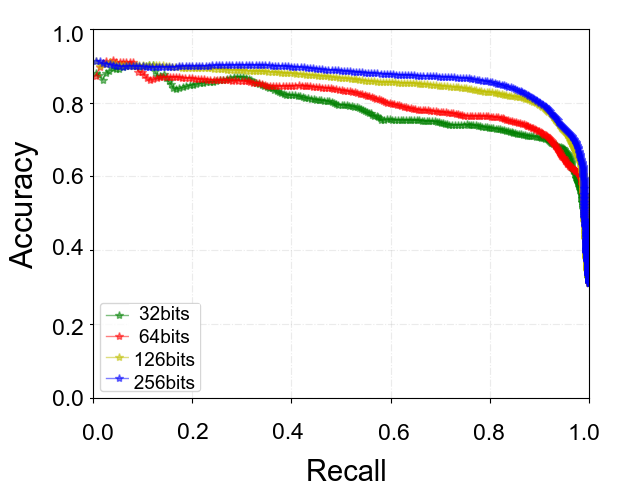}
    \caption{Accuracy-recall curves of different code lengths.}
    \label{fig:fig8}
\end{figure}

We plot the Accuracy-Recall curves (Fig. \ref{fig:fig8}) to evaluate the performance of the model setting the code length to be 32, 64, 128, and 256 bits. Each value of ``Accuracy'' and ``Recall'' in the Fig. \ref{fig:fig8} is the average value of ten thousands ANNS results. We perform ten thousand times of ANNSs to guarantee each of 21,737 samples can be selected. Therefore, a unbiased statistical result is obtained. The training set and validation set of each ANNS is randomly selected. In Fig. \ref{fig:fig8}, the larger area under the curve suggests the better performance intuitively, that is, both the ``Accuracy'' and ``Recall'' achieve a higher level. It can be observed that as the code length increases, the performance of the model is improved. But to a certain extent, the variation of performance is less sensitive to the code length. Therefore, we ultimately choose 256 as the code length. 

Finally, we obtain an effective ML-PIL hash learning model which had both high ``Accuracy'' and ``Recall''. Besides, the time consumed for training the ML-PIL framework is 11.76s, which is much less than that of the traditional gradient descent based deep learning networks.
 
\subsection{Application of ML-PIL based hash learning to recognize M-type spectra}
\label{sec:application}
\begin{figure}
	\centering
	\includegraphics[width=0.9\columnwidth]{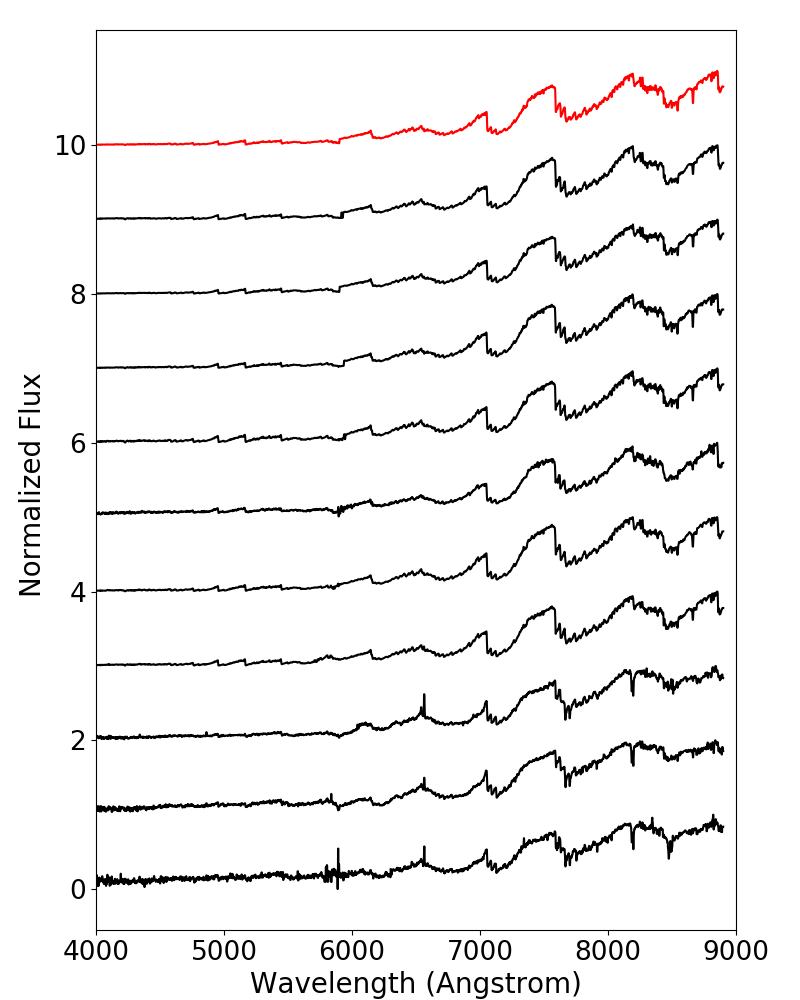}
    \caption{Example of the top 10 search results for a late-type M spectrum (red). The top 10 spectra (black) are sorted by decreasing similarities.}
    \label{fig:fig9}
\end{figure}

\begin{figure*}
	\includegraphics[width=0.99\textwidth]{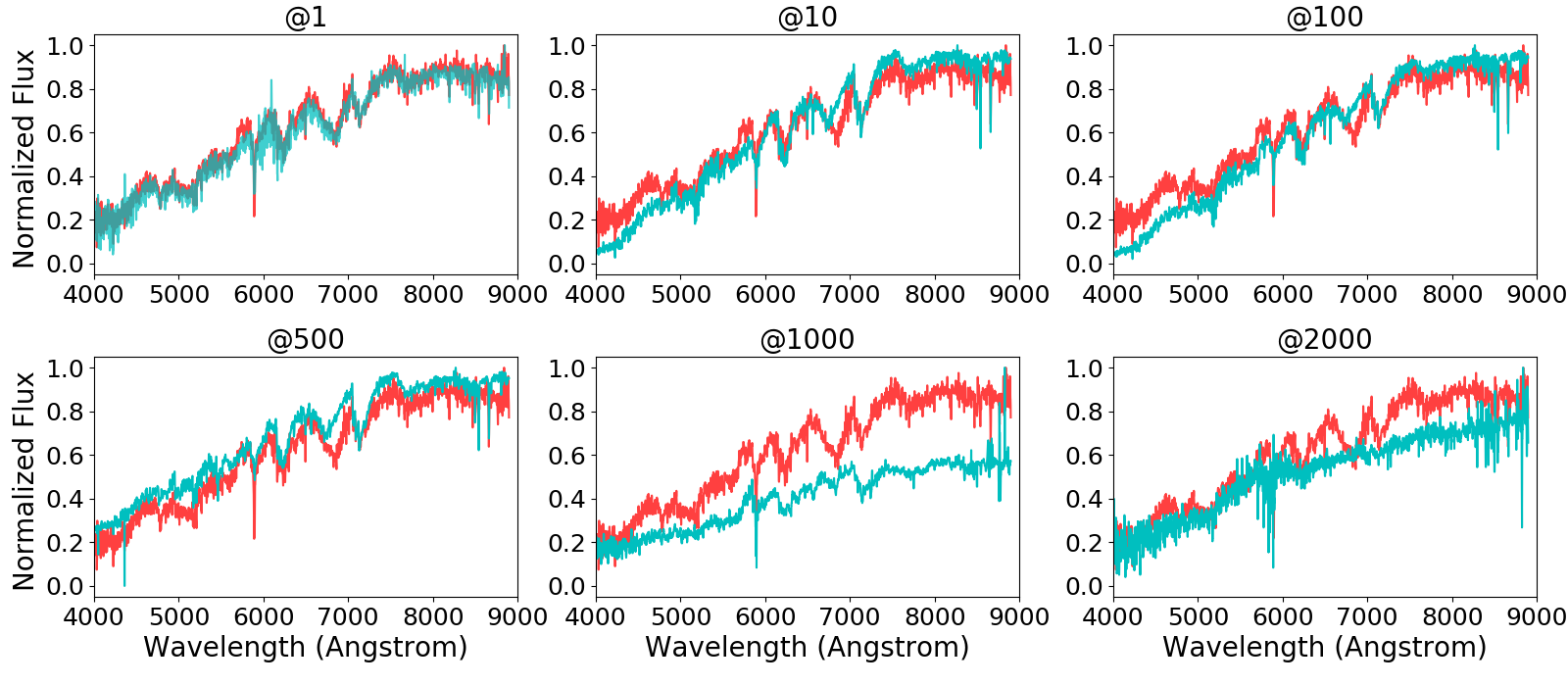}
    \caption{Retrieved spectra (blue) in different positions of the ranked list for an early-type M query (red), where the number after @ denotes the ranking.}
    \label{fig:fig10}
\end{figure*}

We apply the ML-PIL hash learning method to search for M-type spectra from ``UNKNOWN'' data in LAMOST DR5 V1 with the 6,737 query (positive) sample spectra which are described  in subsection \ref{sec:positive}. We firstly derive the hash codes for the query samples and all ``UNKNOWN" spectra through the ML-PIL hash model. Then, for each ``query'' we calculate similarity between the query sample and each ``UNKNOWN'' spectrum using the Hamming distance between their hash codes. The smaller distance the better similarity.  Fig. \ref{fig:fig9} shows one example of the top 10 search results for a late-type M spectrum. On the other hand, Fig. \ref{fig:fig10} shows another example of the increasing dissimilarity with the Hamming distance for an early-type M spectrum. Those similarities ranks top 10\% are kept for each of 6,737 searches. 

\begin{figure}
	\includegraphics[width=\columnwidth]{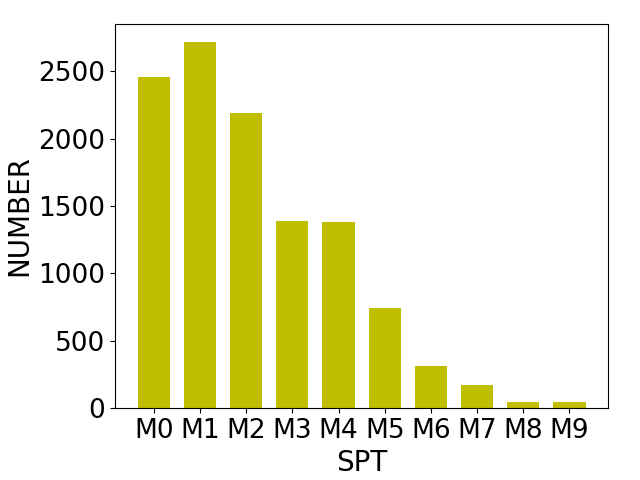}
    \caption{Subtypes distribution of the 11,410 M-type spectra.}
    \label{fig:fig11}
\end{figure}

\begin{figure}
	\includegraphics[width=\columnwidth]{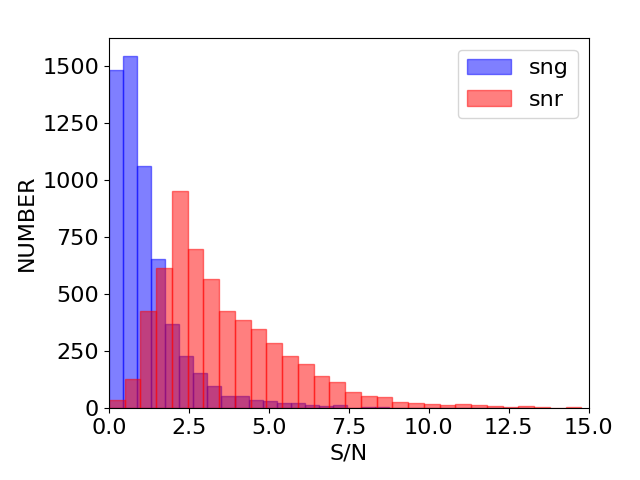}
    \caption{S/Ns distribution of the 11,410 M-type spectra.}
    \label{fig:fig12}
\end{figure}

We manually inspect the union of these 6,737 subset and recognized  11,410 M-type spectra (11,156 objects) including 10,242 dwarf and 1,168 spectra from the 642,178 ``UNKNOWN'' spectra in LAMOST DR5 V1. We make a supplemental  catalog and re-archive all these 11,410 spectra from ``UNKNOWN" category into M-type star in LAMOST DR5 V2, which will be officially released in June 2019. Like former LAMOST data releases, we measure same parameters for these spectra in the catalog including indices of nine molecular bands, equivalent width of H$\alpha$, magnetic activity, and metal-sensitive parameter $\zeta$ \citep{2014AJ....147...33Y,2015RAA....15.1182G} etc. In addition, the catalog also provides spectral subtype for these spectra determined using an improved Hammer package.  The improvement to the original Hammer \citep{2007AJ....134.2398C} was made by \cite{2014AJ....147...33Y} who incorporated three new indices to increase the classification correctness. In the catalog, each object also has radial velocity which is measured through cross-matching with dwarf templates, and the giant/dwarf separation which is determined using the suggested methods described in Section \ref{indicator}, respectively. This supplemental catalog can be downloaded from the web site \url{http://paperdata.china-vo.org/Guoyx/2018/M_etable.txt}. Table ~\ref{tab:table2} shows the first five rows of the catalog. Fig. \ref{fig:fig11} and Fig. \ref{fig:fig12} show the distributions of the spectral subtypes and the S/Ns of the 11,410 spectra, respectively.

\begin{figure}
	\includegraphics[width=\columnwidth]{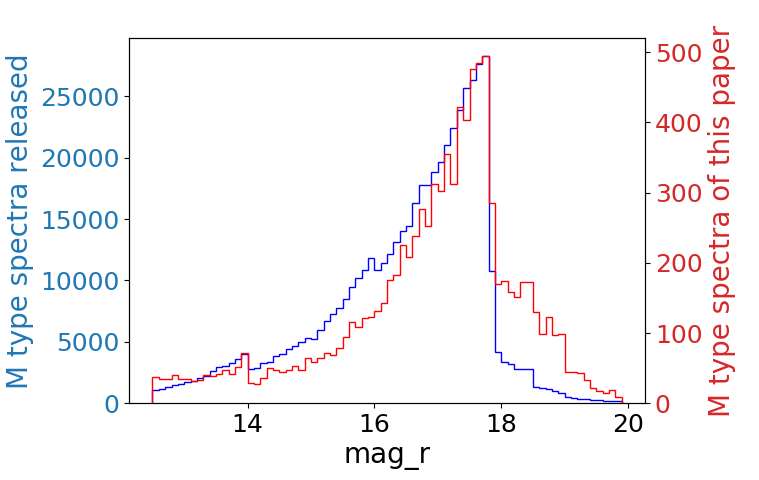}
    \caption{Comparison of magnitude (in r band) distribution of M-type spectra between released in LAMOST DR5 V1and recognized through ML-PIL hash learning. The magnitude distribution of M-type spectra in DR5 V1 are shown in blue with the left vertical axis while that from ML-PIL hash learning are in red with the right vertical axis.}
    \label{fig:fig13}
\end{figure}

\begin{figure}
        \includegraphics[width=\columnwidth]{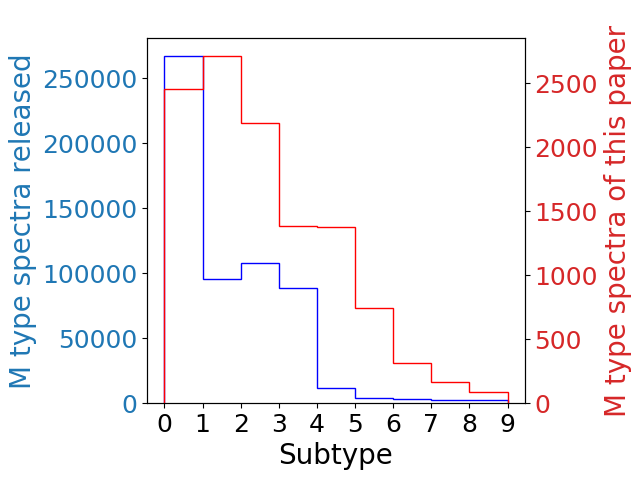}
      \caption{Comparison of subtype distribution of M-type spectra between released in LAMOST DR5 V1 and recognized through ML-PIL hash learning.  The subtype distribution of M-type spectra in DR5 V1 are shown in blue with the left vertical axis while that from ML-PIL hash learning are in red with the right vertical axis.}
    \label{fig:fig14}
\end{figure}

The number distributions of the M-type spectra in LAMOST DR5 V1 and the supplemental spectra in mag\_r space are compared and shown in Fig. \ref{fig:fig13}. These supplemental M-type spectra not only have fainter luminosity, but also have higher proportion of the late-type than the M-type spectra in LAMOST DR5 V1, and the comparison are shown in Fig. \ref{fig:fig14}.   The total number of late M-type spectra (later than M5) recognized through ML-PIL based hash learning is 569.

Adding 11,410 M-type spectra from ``UNKNOWN'' data in LAMOST DR5 V1 to the M dataset of  LAMOST DR5 V1 which originally has 58,3728 M-type spectra, we now posses a larger M star catalog for DR5 (defined as ``ALL M'' hereafter) to study the giant/dwarf separation and the magnetic activity in the discussion section.

\section{Discussion}
\label{sec:discussion}

\subsection{Luminosity class indicators}
\label{indicator}
\begin{figure}
	\centering
	\subfigure{\includegraphics[width=0.8\columnwidth]{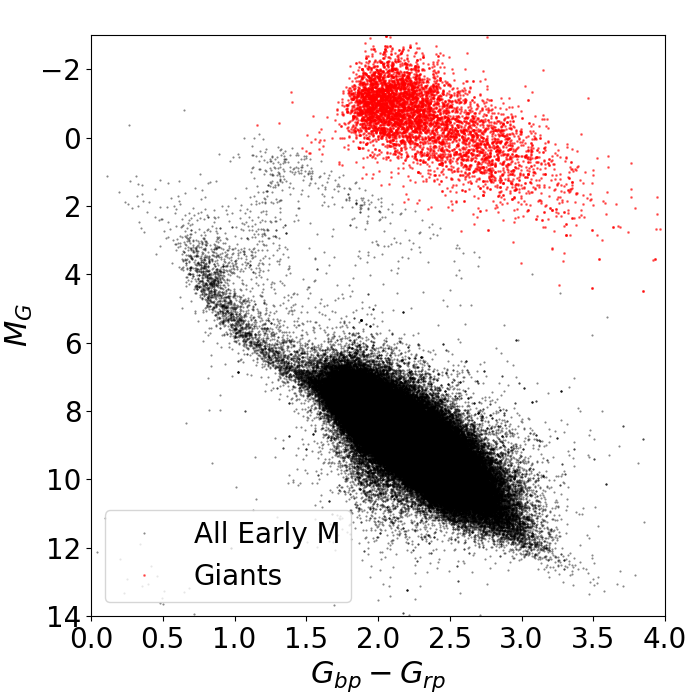}}
	\subfigure{\includegraphics[width=0.8\columnwidth]{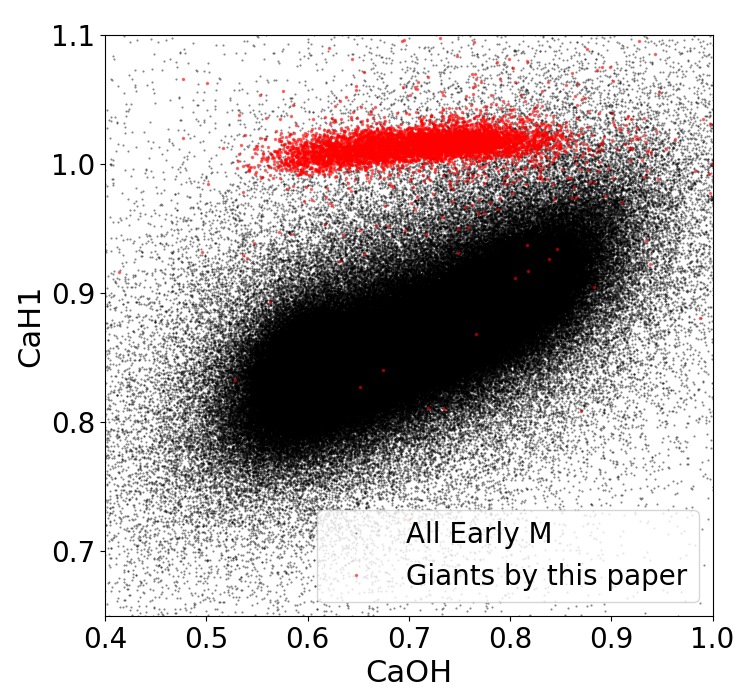}}
	\subfigure{\includegraphics[width=0.8\columnwidth]{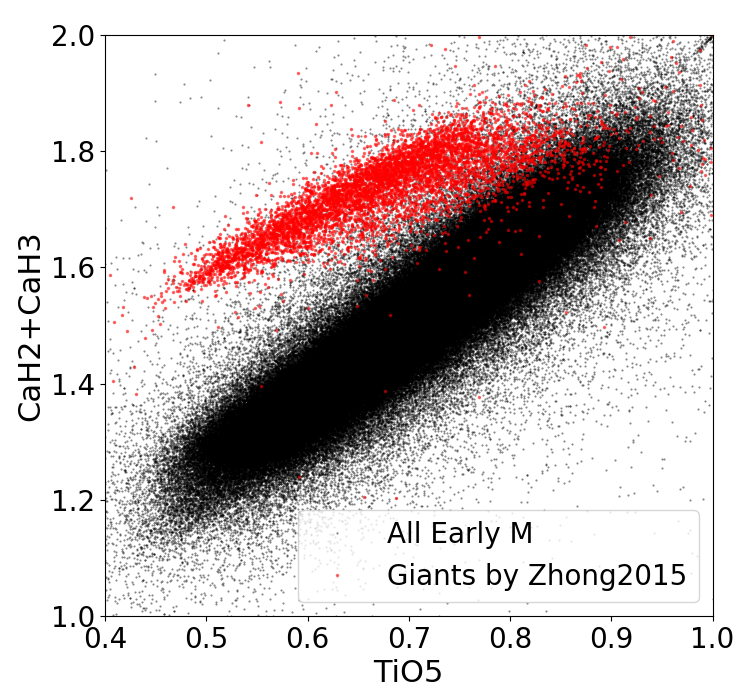}}
    \caption{ The distribution of early M-type stars in the $Gaia$ HRD (Top panel), CaH1 versus CaOH (middle panel) and CaH2+CaH3 versus TiO5 (bottom panel) diagrams. Black and red dots denote the dwarfs and giants respectively. For the same subsample of upper branch in the $Gaia$ HRD, a more clear separation can be seen in the CaH1 versus CaOH diagram, whereas a small portion of giants are mixed into the dwarf branch in the CaH2+CaH3 versus TiO5 diagram.}
    \label{fig:fig15}
\end{figure}

\begin{figure}
	\centering
	\subfigure{\includegraphics[width=0.8\columnwidth]{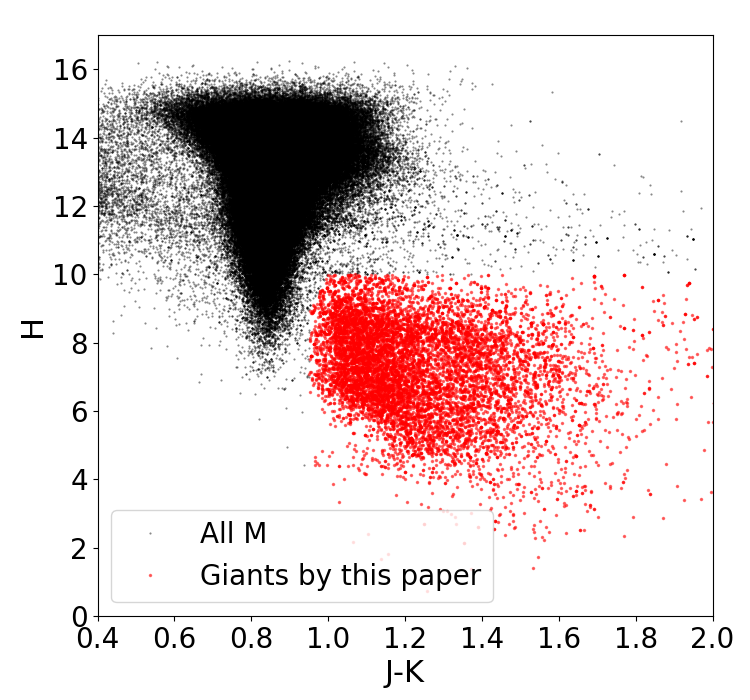}}
	\subfigure{\includegraphics[width=0.8\columnwidth]{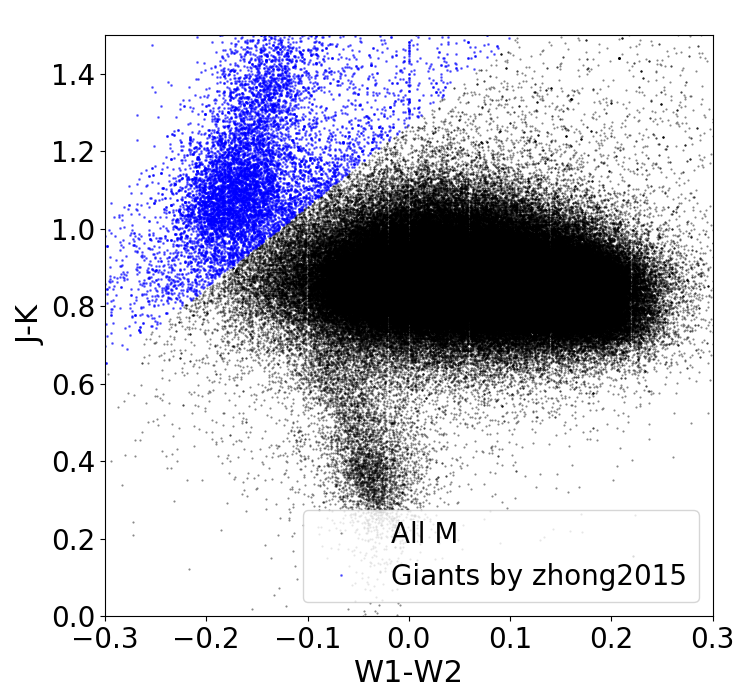}}
	\subfigure{\includegraphics[width=0.8\columnwidth]{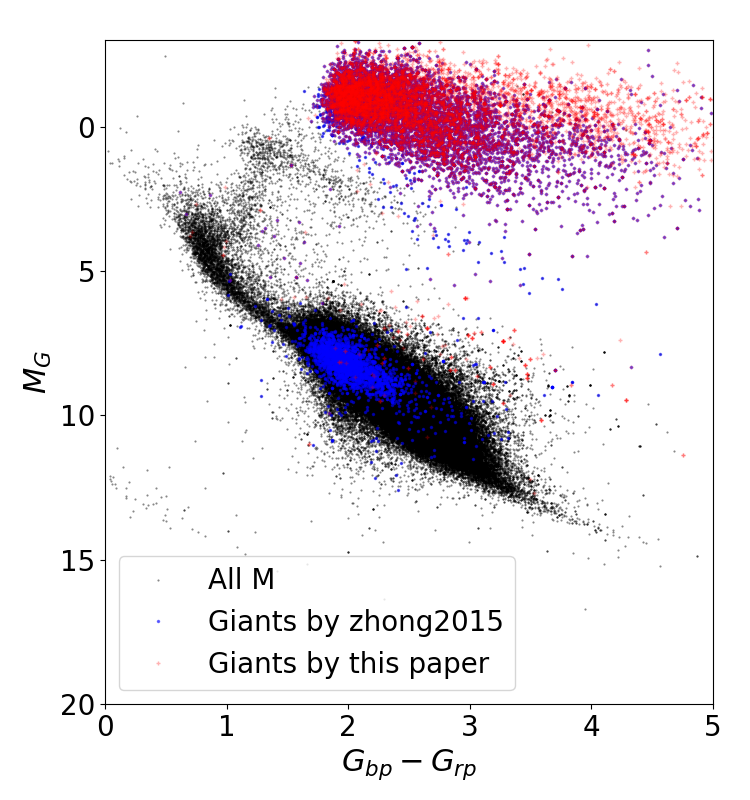}}
    \caption{ The distribution of ``ALL M'' stars in the $H$ versus $J-K$ (Top panel), $J-K$ versus $W1-W2$ (middle panel) and $Gaia$ HRD (bottom panel) diagrams.Giants(red dots) determined in the $H$ versus $J-K$ diagram lie in the upper branch of $Gaia$ HRD. While giants(blue dots) determined in the $J-K$ versus $W1-W2$ diagram, lie mainly in the upper strip of the $Gaia$ HRD, with a small portion lying in the main-sequence strip.}
    \label{fig:fig16}
\end{figure}

We use ``ALL M" objects to check both the spectroscopic and the photometric criteria for separation of M giant and dwarf proposed by \cite{2015RAA....15.1154Z} (Zhong2015 for short), which are the CaH2+CaH3 versus TiO5 line index diagram and the $J-K$ versus $W1-W2$ color diagram respectively, and suggest better spectroscopic and photometric separator for M giant/dwarf, which are the CaOH versus CaH1 line index diagram (middle panel of Fig. \ref{fig:fig15}) and the $H$ versus $J-K$ color diagram (top panel of Fig. \ref{fig:fig16}). We use HRD of $Gaia$ DR2 to verify the suggested separation approach.

Accurate parallaxes and proper motions for the vast majority of ``ALL M'' are obtained through cross-matching within 5 arcsec to Gaia DR2 \citep{2018arXiv180409382G} which have come available in April 2018. We build the $Gaia$ HRDs by simply estimating the absolute $Gaia$ magnitude in the G band for individual star using $M_{G} = G+5+5\log_{10}(\varpi/1000)$, with $\varpi$ the parallax in miliarcseconds (plus the extinction) \citep{2018A&A...616A..10G}. This is valid when the relative uncertainty on the parallax is $<\sim$ 20\% \citep{2018A&A...616A...9L}.

First, we choose the early M-type spectra to analyze the validation of the luminosity discrimination in spectral features. As shown in the $Gaia$ HRD, the top panel of Fig. \ref{fig:fig15}, the M giants are in red color and locate in the upper branch while the M dwarfs are clearly separated in black color and locate in the lower branch. The middle panel of Fig. \ref{fig:fig15}, the CaOH versus CaH1 diagram, shows that the same giants population with the upper panel in red color lay in the upper branch in this diagram. However, Zhong2015 was weaker to discriminate M giants and dwarfs for early M type spectra.  The bottom panel of Fig. \ref{fig:fig15}, the CaH2+CaH3 versus TiO5 diagram, shows some giants overlap with dwarfs in the lower branch where is the location of dwarfs. This overlap means that the criterion in Zhong2015 will lead to a small portion of M giants misclassified as dwarfs.

Then, we examine the effectiveness of the $H$ versus $J-K$ criterion using the total ``ALL M'', and we can see different loci of dwarfs and giants in the bottom of $Gaia$ HRD. As shown in top and middle panel of Fig. \ref{fig:fig16}, both the suggested criteria in this paper and Zhong2015 can separate giants and dwarfs. In these two panel, dwarfs are shown in black color, while giants are in red or blue represent classified by the criteria in this paper or by Zhong2015 respectively. Comparing this two groups of giants from different separator in the the $Gaia$ HRD shown in the bottom panel of Fig. \ref{fig:fig16}, part of giant candidates ($\sim$12\%) from the criterion in Zhong2015, $J-K$ versus $W1-W2$, should actually be dwarfs lying in the main-sequence strip. It is clear that  $H$ versus $J-K$ can easier eliminate possible dwarf contaminations from giants than the method given in Zhong2015.  

Using both the spectral features and the 2MASS photometry, we determine each M-type spectra as giant or dwarf in the supplemental catalog. From the result, we conclude that even lacking of 2MASS infrared data we still can efficient to separate M giants from M dwarfs based on spectral feature.

\begin{figure}
	\subfigure{\includegraphics[width=\columnwidth]{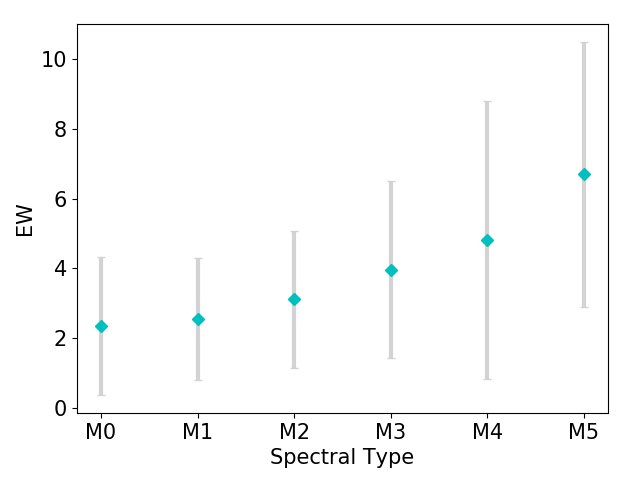}}
	\subfigure{\includegraphics[width=\columnwidth]{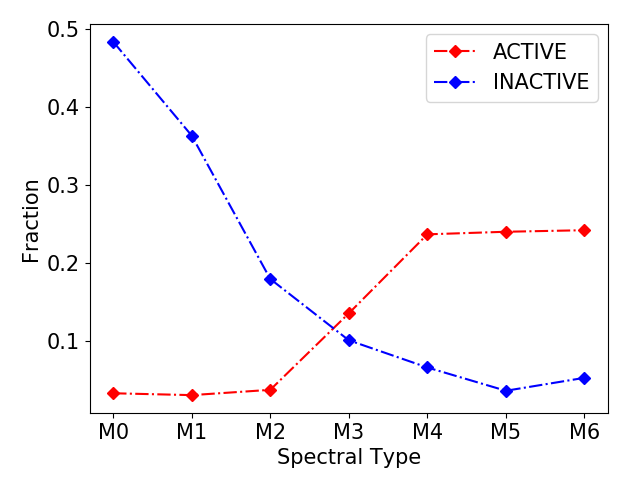}}
    \caption{H$\alpha$ EWs (upper panel) and magnetic activity fraction (lower panel) as a function of spectral type. The red (blue) dots in the lower panel is the ratio of active (inactive) stars.}
    \label{fig:fig17}
\end{figure}

\subsection{Magnetic activity and kinematics}
Magnetic fields affect the chromospheric activity of M dwarfs, and H$\alpha$ emission can be an indicator of chromospheric activity. We investigate the magnetic activity of M dwarfs by measuring the equivalent widths (EWs) of H$\alpha$. Once the S/N of the continuum around H$\alpha$ of a M dwarf is greater than 3, the M dwarf spectra is then to be checked the value of EW of the H$\alpha$ greater or less than 1 to determine it is active or inactive.\citep{2015RAA....15.1182G}. If the S/N around H$\alpha$ of a M dwarf less than 3, the activity of the M dwarf will not be measured. The upper panel of Fig.\ref{fig:fig17} shows that the EWs of H$\alpha$ increase with the subtype becoming later, while the lower panel shows that mean fraction of active stars increases or inactive stars decreases with spectral subtype becoming later. This implies that later M dwarfs show stronger and higher fraction of magnetic activity.

We also investigate the velocities and velocity dispersions for both active and inactive M dwarfs. Combining radial velocities, distances and proper motions from $Gaia$, the heliocentric space motions ($U,V,W$) are computed according to the method of \citep{1987PhRvC..36.2252J}. The 3D velocities are computed in a right-handed coordinate system, with positive $U$ velocity toward the Galactic center, positive $V$ velocity in the direction of Galactic rotation and positive $W$ velocity toward the north Galactic pole. The velocities are corrected for solar motion (10, 5, 7 $\rm km/s^{-1}$) \citep{1998MNRAS.298..387D} with respect to the local standard of rest. These kinematical parameters are also provided in the supplemental catalog. 

\begin{figure*}
	\includegraphics[width=\textwidth]{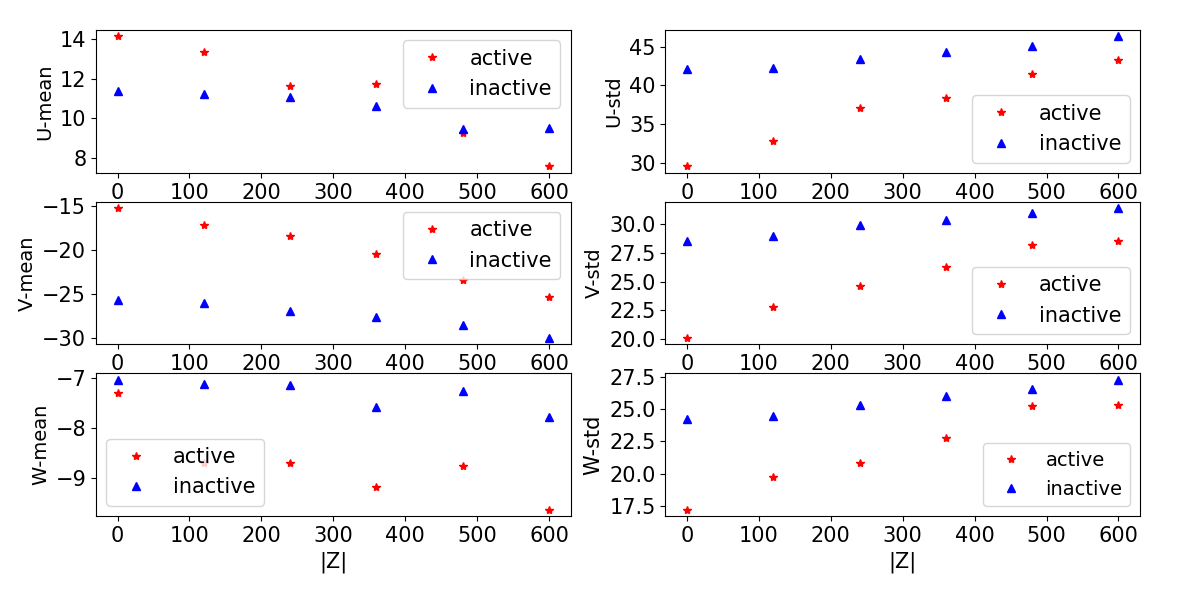}
    \caption{\textbf{$UVW$ velocity mean values (left) and velocity dispersions (right) as a function of absolute vertical distance from the Galactic plane in 100 pc bins for active (red asterisks) and inactive (blue triangles) M dwarfs. }}
    \label{fig:fig18}
\end{figure*}

The M dwarfs are binned in 100 pc increments of absolute vertical distance from the Galactic plane. The $UVW$ velocity mean values and velocity dispersions as a function of absolute vertical distance for active and inactive populations are shown in Fig.\ref{fig:fig18}. From the figure, we can see that the active M dwarfs are systematically low in velocity dispersion in the $W$ direction. While the the velocity mean values of the active M dwarfs are high in $U$ and $V$ directions. The two populations separated apparently, suggesting that the active M dwarfs should be born in an older kinematical population, which is consistent with \cite{2011ASPC..448.1359H}. The $UVW$ velocity mean values decline with increasing absolute vertical distance, whereas the $UVW$ velocity dispersions rise, for both the active and inactive populations. This result agrees well with the trend for thin disks shown in \cite{2007AJ....134.2418B}. 

Furthermore, although we find that the the strength of H$\alpha$ emission line varies in multiple observations for some M dwarfs, we don't have enough data to draw any conclusion, which needs analysis of other physical characteristics, such as flare, rotation, and their intrinsic relationships by using time domain photometric and spectroscopic observations.

\section{SUMMARY}

A binary nonlinear hashing algorithm based on ML-PIL is proposed to effectively learn spectral features of M-type stars, in order to search for missing M type stars due to failures of multi-template matching particularly for low signal-to-noise ratio spectra. We construct a specific ML-PIL model for the learning and searching, and build a positive sample through clustering both high and low S/N known M-type spectra. Evaluating the performance of the model and effectively applying to 642,178 ``UNKNOWN" spectra in LAMOST DR5 V1, we finally recognize 11,410 M-type spectra and make a catalog to supplement to the M-type star catalog of LAMOST DR5 V1. For the recognized spectra, some useful values are calculated including indices of molecular bands, magnetic activities and metal-sensitive parameters $\zeta$. Adding the M-type spectra recognized through ML-PIL to the original released M-type stars in DR5 V1, we obtain a complete catalog of M-type stars in LAMOST DR5 which will be officially released in June 2019. Through cross-matching, the common objects with $Gaia$ DR2 are used to study the giant/dwarf separators based on the 2MASS color indices and and LAMOST spectral line indices. We then propose two giant/dwarf separators, and verify them with the HRD from $Gaia$ DR2, by which we label the objects as dwarfs or giants and calculate kinematics for the M dwarfs. According to the good performance of ML-PIL based hash learning algorithm and their successful application in M-type spectra search, we believe it is able to effectively search for specific spectra, especially low S/N data such as LAMOST ``UNKNOWN" dataset in which there still potentially exist early type stars besides M-type stars.

\section*{ACKNOWLEDGEMENTS}

This research is supported by the Major State Basic Research Development Program of China (973 Program, No. 2014CB845700), China Scholarship Council and the National Natural Science Foundation of China (Grant Nos. 11703053 and 11703051). This research has made use of LAMOST data. The Guo Shou Jing Telescope (the Large Sky Area Multi-Object Fiber Spectroscopic Telescope, LAMOST) is a National Major Scientific Project built by the Chinese Academy of Sciences. Funding for the project has been provided by the National Development and Reform Commission. This research also makes use of data products from the Two Micron All Sky Survey, which is a joint project of the University of Massachusetts and the Infrared Processing and Analysis Center/California Institute of Technology, funded by the National Aeronautics and Space Administration and the National Science Foundation.




\bibliographystyle{mnras}
\bibliography{article.bib} 

\begin{thebibliography}{}
\makeatletter
\relax
\def\mn@urlcharsother{\let\do\@makeother \do\$\do\&\do\#\do\^\do\_\do\%\do\~}
\def\mn@doi{\begingroup\mn@urlcharsother \@ifnextchar [ {\mn@doi@}
  {\mn@doi@[]}}
\def\mn@doi@[#1]#2{\def\@tempa{#1}\ifx\@tempa\@empty \href
  {http://dx.doi.org/#2} {doi:#2}\else \href {http://dx.doi.org/#2} {#1}\fi
  \endgroup}
\def\mn@eprint#1#2{\mn@eprint@#1:#2::\@nil}
\def\mn@eprint@arXiv#1{\href {http://arxiv.org/abs/#1} {{\tt arXiv:#1}}}
\def\mn@eprint@dblp#1{\href {http://dblp.uni-trier.de/rec/bibtex/#1.xml}
  {dblp:#1}}
\def\mn@eprint@#1:#2:#3:#4\@nil{\def\@tempa {#1}\def\@tempb {#2}\def\@tempc
  {#3}\ifx \@tempc \@empty \let \@tempc \@tempb \let \@tempb \@tempa \fi \ifx
  \@tempb \@empty \def\@tempb {arXiv}\fi \@ifundefined
  {mn@eprint@\@tempb}{\@tempb:\@tempc}{\expandafter \expandafter \csname
  mn@eprint@\@tempb\endcsname \expandafter{\@tempc}}}

\bibitem[\protect\citeauthoryear{Almeida \& Prieto}{Almeida \&
  Prieto}{2013}]{0004-637X-763-1-50}
Almeida J.~S.,  Prieto C.~A.,  2013, The Astrophysical Journal, 763, 50

\bibitem[\protect\citeauthoryear{Andoni \& Indyk}{Andoni \&
  Indyk}{2006}]{Andoni2006Near}
Andoni A.,  Indyk P.,  2006, in IEEE Symposium on Foundations of Computer
  Science. pp 459--468

\bibitem[\protect\citeauthoryear{Arthur \& Vassilvitskii}{Arthur \&
  Vassilvitskii}{2007}]{Arthur2007k}
Arthur D.,  Vassilvitskii S.,  2007, in Eighteenth Acm-Siam Symposium on
  Discrete Algorithms, New Orleans, Louisiana. pp 1027--1035

\bibitem[\protect\citeauthoryear{{Bayo} et~al.,}{{Bayo}
  et~al.}{2017}]{2017MNRAS.465..760B}
{Bayo} A.,  et~al., 2017, \mn@doi [\mnras] {10.1093/mnras/stw2760}, \href
  {http://adsabs.harvard.edu/abs/2017MNRAS.465..760B} {465, 760}

\bibitem[\protect\citeauthoryear{{Benedict} et~al.,}{{Benedict}
  et~al.}{2016}]{2016AJ....152..141B}
{Benedict} G.~F.,  et~al., 2016, \mn@doi [\aj] {10.3847/0004-6256/152/5/141},
  \href {http://adsabs.harvard.edu/abs/2016AJ....152..141B} {152, 141}

\bibitem[\protect\citeauthoryear{{Bessell} \& {Brett}}{{Bessell} \&
  {Brett}}{1988}]{1988PASP..100.1134B}
{Bessell} M.~S.,  {Brett} J.~M.,  1988, \mn@doi [\pasp] {10.1086/132281}, \href
  {http://adsabs.harvard.edu/abs/1988PASP..100.1134B} {100, 1134}

\bibitem[\protect\citeauthoryear{{Bessell}, {Castelli}  \& {Plez}}{{Bessell}
  et~al.}{1998}]{1998A&A...337..321B}
{Bessell} M.~S.,  {Castelli} F.,   {Plez} B.,  1998, \aap, \href
  {http://adsabs.harvard.edu/abs/1998A%26A...337..321B} {337, 321}

\bibitem[\protect\citeauthoryear{{Bochanski}, {Munn}, {Hawley}, {West}, {Covey}
   \& {Schneider}}{{Bochanski} et~al.}{2007}]{2007AJ....134.2418B}
{Bochanski} J.~J.,  {Munn} J.~A.,  {Hawley} S.~L.,  {West} A.~A.,  {Covey}
  K.~R.,   {Schneider} D.~P.,  2007, \mn@doi [\aj] {10.1086/522053}, \href
  {http://adsabs.harvard.edu/abs/2007AJ....134.2418B} {134, 2418}

\bibitem[\protect\citeauthoryear{{Bochanski}, {Hawley}, {Covey}, {West},
  {Reid}, {Golimowski}  \& {Ivezi{\'c}}}{{Bochanski}
  et~al.}{2010}]{2010AJ....139.2679B}
{Bochanski} J.~J.,  {Hawley} S.~L.,  {Covey} K.~R.,  {West} A.~A.,  {Reid}
  I.~N.,  {Golimowski} D.~A.,   {Ivezi{\'c}} {\v Z}.,  2010, \mn@doi [\aj]
  {10.1088/0004-6256/139/6/2679}, \href
  {http://adsabs.harvard.edu/abs/2010AJ....139.2679B} {139, 2679}

\bibitem[\protect\citeauthoryear{Bondugula}{Bondugula}{2013}]{Bondugula_surveyof}
Bondugula S.,  2013, Survey of Hashing Techniques for Compact Bit
  Representations of Images

\bibitem[\protect\citeauthoryear{{Chabrier}}{{Chabrier}}{2003}]{2003PASP..115..763C}
{Chabrier} G.,  2003, \mn@doi [\pasp] {10.1086/376392}, \href
  {http://adsabs.harvard.edu/abs/2003PASP..115..763C} {115, 763}

\bibitem[\protect\citeauthoryear{{Chabrier}, {Baraffe}, {Allard}  \&
  {Hauschildt}}{{Chabrier} et~al.}{2000}]{2000ApJ...542..464C}
{Chabrier} G.,  {Baraffe} I.,  {Allard} F.,   {Hauschildt} P.,  2000, \mn@doi
  [\apj] {10.1086/309513}, \href
  {http://adsabs.harvard.edu/abs/2000ApJ...542..464C} {542, 464}

\bibitem[\protect\citeauthoryear{{Covey} et~al.,}{{Covey}
  et~al.}{2007}]{2007AJ....134.2398C}
{Covey} K.~R.,  et~al., 2007, \mn@doi [\aj] {10.1086/522052}, \href
  {http://adsabs.harvard.edu/abs/2007AJ....134.2398C} {134, 2398}

\bibitem[\protect\citeauthoryear{{Covey} et~al.,}{{Covey}
  et~al.}{2008}]{2008AJ....136.1778C}
{Covey} K.~R.,  et~al., 2008, \mn@doi [\aj] {10.1088/0004-6256/136/5/1778},
  \href {http://adsabs.harvard.edu/abs/2008AJ....136.1778C} {136, 1778}

\bibitem[\protect\citeauthoryear{{Cui} et~al.,}{{Cui}
  et~al.}{2012}]{2012RAA....12.1197C}
{Cui} X.-Q.,  et~al., 2012, \mn@doi [Research in Astronomy and Astrophysics]
  {10.1088/1674-4527/12/9/003}, \href
  {http://adsabs.harvard.edu/abs/2012RAA....12.1197C} {12, 1197}

\bibitem[\protect\citeauthoryear{{Dehnen} \& {Binney}}{{Dehnen} \&
  {Binney}}{1998}]{1998MNRAS.298..387D}
{Dehnen} W.,  {Binney} J.~J.,  1998, \mn@doi [\mnras]
  {10.1046/j.1365-8711.1998.01600.x}, \href
  {http://adsabs.harvard.edu/abs/1998MNRAS.298..387D} {298, 387}

\bibitem[\protect\citeauthoryear{{Delfosse}, {Forveille}, {S{\'e}gransan},
  {Beuzit}, {Udry}, {Perrier}  \& {Mayor}}{{Delfosse}
  et~al.}{2000}]{2000A&A...364..217D}
{Delfosse} X.,  {Forveille} T.,  {S{\'e}gransan} D.,  {Beuzit} J.-L.,  {Udry}
  S.,  {Perrier} C.,   {Mayor} M.,  2000, \aap, \href
  {http://adsabs.harvard.edu/abs/2000A%26A...364..217D} {364, 217}

\bibitem[\protect\citeauthoryear{{Feiden} \& {Chaboyer}}{{Feiden} \&
  {Chaboyer}}{2014}]{2014ApJ...789...53F}
{Feiden} G.~A.,  {Chaboyer} B.,  2014, \mn@doi [\apj]
  {10.1088/0004-637X/789/1/53}, \href
  {http://adsabs.harvard.edu/abs/2014ApJ...789...53F} {789, 53}

\bibitem[\protect\citeauthoryear{{Gaia Collaboration} et~al.,}{{Gaia
  Collaboration} et~al.}{2018a}]{2018arXiv180409382G}
{Gaia Collaboration} et~al., 2018a, preprint, \href
  {http://adsabs.harvard.edu/abs/2018arXiv180409382G} {} (\mn@eprint {arXiv}
  {1804.09382})

\bibitem[\protect\citeauthoryear{{Gaia Collaboration} et~al.,}{{Gaia
  Collaboration} et~al.}{2018b}]{2018A&A...616A..10G}
{Gaia Collaboration} et~al., 2018b, \mn@doi [\aap]
  {10.1051/0004-6361/201832843}, \href
  {http://adsabs.harvard.edu/abs/2018A%26A...616A..10G} {616, A10}

\bibitem[\protect\citeauthoryear{Gong \& Lazebnik}{Gong \&
  Lazebnik}{2011}]{Gong2011Iterative}
Gong Y.,  Lazebnik S.,  2011, in Computer Vision and Pattern Recognition. pp
  817--824

\bibitem[\protect\citeauthoryear{{Guo} et~al.,}{{Guo}
  et~al.}{2015}]{2015RAA....15.1182G}
{Guo} Y.-X.,  et~al., 2015, \mn@doi [Research in Astronomy and Astrophysics]
  {10.1088/1674-4527/15/8/007}, \href
  {http://adsabs.harvard.edu/abs/2015RAA....15.1182G} {15, 1182}

\bibitem[\protect\citeauthoryear{Guo, Wang  \& Xin}{Guo
  et~al.}{2017}]{Guo2017Autoencoder}
Guo P.,  Wang K.,   Xin X.,  2017, in Smc.

\bibitem[\protect\citeauthoryear{{Han} et~al.,}{{Han}
  et~al.}{2017}]{2017AJ....154..100H}
{Han} E.,  et~al., 2017, \mn@doi [\aj] {10.3847/1538-3881/aa803c}, \href
  {http://adsabs.harvard.edu/abs/2017AJ....154..100H} {154, 100}

\bibitem[\protect\citeauthoryear{{Hawley}, {Bochanski}  \& {West}}{{Hawley}
  et~al.}{2011}]{2011ASPC..448.1359H}
{Hawley} S.~L.,  {Bochanski} J.~J.,   {West} A.~A.,  2011, in {Johns-Krull} C.,
   {Browning} M.~K.,   {West} A.~A.,  eds,  Astronomical Society of the Pacific
  Conference Series Vol. 448, 16th Cambridge Workshop on Cool Stars, Stellar
  Systems, and the Sun. p.~1359 (\mn@eprint {arXiv} {1012.3505})

\bibitem[\protect\citeauthoryear{{Henry} \& {McCarthy}}{{Henry} \&
  {McCarthy}}{1993}]{1993AJ....106..773H}
{Henry} T.~J.,  {McCarthy} Jr. D.~W.,  1993, \mn@doi [\aj] {10.1086/116685},
  \href {http://adsabs.harvard.edu/abs/1993AJ....106..773H} {106, 773}

\bibitem[\protect\citeauthoryear{{Henry}, {Kirkpatrick}  \& {Simons}}{{Henry}
  et~al.}{1994}]{1994AJ....108.1437H}
{Henry} T.~J.,  {Kirkpatrick} J.~D.,   {Simons} D.~A.,  1994, \mn@doi [\aj]
  {10.1086/117167}, \href {http://adsabs.harvard.edu/abs/1994AJ....108.1437H}
  {108, 1437}

\bibitem[\protect\citeauthoryear{{Henry}, {Jao}, {Subasavage}, {Beaulieu},
  {Ianna}, {Costa}  \& {M{\'e}ndez}}{{Henry}
  et~al.}{2006}]{2006AJ....132.2360H}
{Henry} T.~J.,  {Jao} W.-C.,  {Subasavage} J.~P.,  {Beaulieu} T.~D.,  {Ianna}
  P.~A.,  {Costa} E.,   {M{\'e}ndez} R.~A.,  2006, \mn@doi [\aj]
  {10.1086/508233}, \href {http://adsabs.harvard.edu/abs/2006AJ....132.2360H}
  {132, 2360}

\bibitem[\protect\citeauthoryear{Heo, Lee, He, Chang  \& Yoon}{Heo
  et~al.}{2012}]{Heo2012Spherical}
Heo J.~P.,  Lee Y.,  He J.,  Chang S.~F.,   Yoon S.~E.,  2012, IEEE, 157, 2957

\bibitem[\protect\citeauthoryear{{Houdebine}, {Mullan}, {Bercu}, {Paletou}  \&
  {Gebran}}{{Houdebine} et~al.}{2017}]{2017ApJ...837...96H}
{Houdebine} E.~R.,  {Mullan} D.~J.,  {Bercu} B.,  {Paletou} F.,   {Gebran} M.,
  2017, \mn@doi [\apj] {10.3847/1538-4357/aa5cad}, \href
  {http://adsabs.harvard.edu/abs/2017ApJ...837...96H} {837, 96}

\bibitem[\protect\citeauthoryear{{Huo} et~al.,}{{Huo}
  et~al.}{2017}]{2017RAA....17...32H}
{Huo} Z.-Y.,  et~al., 2017, \mn@doi [Research in Astronomy and Astrophysics]
  {10.1088/1674-4527/17/4/32}, \href
  {http://adsabs.harvard.edu/abs/2017RAA....17...32H} {17, 032}

\bibitem[\protect\citeauthoryear{{Jackson} \& {Jeffries}}{{Jackson} \&
  {Jeffries}}{2014}]{2014MNRAS.441.2111J}
{Jackson} R.~J.,  {Jeffries} R.~D.,  2014, \mn@doi [\mnras]
  {10.1093/mnras/stu651}, \href
  {http://adsabs.harvard.edu/abs/2014MNRAS.441.2111J} {441, 2111}

\bibitem[\protect\citeauthoryear{{Johnson}, {Horen}  \& {Mahaux}}{{Johnson}
  et~al.}{1987}]{1987PhRvC..36.2252J}
{Johnson} C.~H.,  {Horen} D.~J.,   {Mahaux} C.,  1987, \mn@doi [\prc]
  {10.1103/PhysRevC.36.2252}, \href
  {http://adsabs.harvard.edu/abs/1987PhRvC..36.2252J} {36, 2252}

\bibitem[\protect\citeauthoryear{Jolliffe}{Jolliffe}{2002}]{Jolliffe2002Principal}
Jolliffe I.~T.,  2002, Weather, 98

\bibitem[\protect\citeauthoryear{{L{\'e}pine} \& {Gaidos}}{{L{\'e}pine} \&
  {Gaidos}}{2011}]{2011AJ....142..138L}
{L{\'e}pine} S.,  {Gaidos} E.,  2011, \mn@doi [\aj]
  {10.1088/0004-6256/142/4/138}, \href
  {http://adsabs.harvard.edu/abs/2011AJ....142..138L} {142, 138}

\bibitem[\protect\citeauthoryear{{Li} et~al.,}{{Li}
  et~al.}{2018}]{2018ApJS..234...31L}
{Li} Y.-B.,  et~al., 2018, \mn@doi [\apjs] {10.3847/1538-4365/aaa415}, \href
  {http://adsabs.harvard.edu/abs/2018ApJS..234...31L} {234, 31}

\bibitem[\protect\citeauthoryear{{Luo} et~al.,}{{Luo}
  et~al.}{2015}]{2015RAA....15.1095L}
{Luo} A.-L.,  et~al., 2015, \mn@doi [Research in Astronomy and Astrophysics]
  {10.1088/1674-4527/15/8/002}, \href
  {http://cdsads.u-strasbg.fr/abs/2015RAA....15.1095L} {15, 1095}

\bibitem[\protect\citeauthoryear{{Luri} et~al.,}{{Luri}
  et~al.}{2018}]{2018A&A...616A...9L}
{Luri} X.,  et~al., 2018, \mn@doi [\aap] {10.1051/0004-6361/201832964}, \href
  {http://adsabs.harvard.edu/abs/2018A%26A...616A...9L} {616, A9}

\bibitem[\protect\citeauthoryear{{Mann}, {Gaidos}, {L{\'e}pine}  \&
  {Hilton}}{{Mann} et~al.}{2012}]{2012ApJ...753...90M}
{Mann} A.~W.,  {Gaidos} E.,  {L{\'e}pine} S.,   {Hilton} E.~J.,  2012, \mn@doi
  [\apj] {10.1088/0004-637X/753/1/90}, \href
  {http://adsabs.harvard.edu/abs/2012ApJ...753...90M} {753, 90}

\bibitem[\protect\citeauthoryear{{Newberg} et~al.,}{{Newberg}
  et~al.}{2012}]{2012ASPC..458..405N}
{Newberg} H.~J.,  et~al., 2012, in {Aoki} W.,  {Ishigaki} M.,  {Suda} T.,
  {Tsujimoto} T.,   {Arimoto} N.,  eds,  Astronomical Society of the Pacific
  Conference Series Vol. 458, Galactic Archaeology: Near-Field Cosmology and
  the Formation of the Milky Way. p.~405

\bibitem[\protect\citeauthoryear{Pal, Hagiwara, Kayaba  \& Morishita}{Pal
  et~al.}{2015}]{Pal2015A}
Pal C.,  Hagiwara I.,  Kayaba N.,   Morishita S.,  2015, Shock \& Vibration, 3,
  201

\bibitem[\protect\citeauthoryear{{Reid}, {Hawley}  \& {Gizis}}{{Reid}
  et~al.}{1995}]{1995AJ....110.1838R}
{Reid} I.~N.,  {Hawley} S.~L.,   {Gizis} J.~E.,  1995, \mn@doi [\aj]
  {10.1086/117655}, \href {http://adsabs.harvard.edu/abs/1995AJ....110.1838R}
  {110, 1838}

\bibitem[\protect\citeauthoryear{{Reiners}}{{Reiners}}{2012}]{2012LRSP....9....1R}
{Reiners} A.,  2012, \mn@doi [Living Reviews in Solar Physics]
  {10.12942/lrsp-2012-1}, \href
  {http://adsabs.harvard.edu/abs/2012LRSP....9....1R} {9, 1}

\bibitem[\protect\citeauthoryear{{Ren}, {Rebassa-Mansergas}, {Parsons}, {Liu},
  {Luo}, {Kong}  \& {Zhang}}{{Ren} et~al.}{2018}]{2018MNRAS.477.4641R}
{Ren} J.-J.,  {Rebassa-Mansergas} A.,  {Parsons} S.~G.,  {Liu} X.-W.,  {Luo}
  A.-L.,  {Kong} X.,   {Zhang} H.-T.,  2018, \mn@doi [\mnras]
  {10.1093/mnras/sty805}, \href
  {http://adsabs.harvard.edu/abs/2018MNRAS.477.4641R} {477, 4641}

\bibitem[\protect\citeauthoryear{Salakhutdinov \& Hinton}{Salakhutdinov \&
  Hinton}{2009}]{Salakhutdinov2009Semantic}
Salakhutdinov R.,  Hinton G.,  2009, International Journal of Approximate
  Reasoning, 50, 969

\bibitem[\protect\citeauthoryear{{Salpeter} \& {Hoffman}}{{Salpeter} \&
  {Hoffman}}{1995}]{1995ApJ...441...51S}
{Salpeter} E.~E.,  {Hoffman} G.~L.,  1995, \mn@doi [\apj] {10.1086/175334},
  \href {http://adsabs.harvard.edu/abs/1995ApJ...441...51S} {441, 51}

\bibitem[\protect\citeauthoryear{{Stassun} et~al.,}{{Stassun}
  et~al.}{2011}]{2011ASPC..448..505S}
{Stassun} K.~G.,  et~al., 2011, in {Johns-Krull} C.,  {Browning} M.~K.,
  {West} A.~A.,  eds,  Astronomical Society of the Pacific Conference Series
  Vol. 448, 16th Cambridge Workshop on Cool Stars, Stellar Systems, and the
  Sun. p.~505 (\mn@eprint {arXiv} {1012.2580})

\bibitem[\protect\citeauthoryear{{Torres}, {Andersen}  \&
  {Gim{\'e}nez}}{{Torres} et~al.}{2010}]{2010A&ARv..18...67T}
{Torres} G.,  {Andersen} J.,   {Gim{\'e}nez} A.,  2010, \mn@doi [\aapr]
  {10.1007/s00159-009-0025-1}, \href
  {http://adsabs.harvard.edu/abs/2010A%26ARv..18...67T} {18, 67}

\bibitem[\protect\citeauthoryear{{Veyette}, {Muirhead}, {Mann}, {Brewer},
  {Allard}  \& {Homeier}}{{Veyette} et~al.}{2017}]{2017ApJ...851...26V}
{Veyette} M.~J.,  {Muirhead} P.~S.,  {Mann} A.~W.,  {Brewer} J.~M.,  {Allard}
  F.,   {Homeier} D.,  2017, \mn@doi [\apj] {10.3847/1538-4357/aa96aa}, \href
  {http://adsabs.harvard.edu/abs/2017ApJ...851...26V} {851, 26}

\bibitem[\protect\citeauthoryear{Wang, Guo, Yin, Luo  \& Xin}{Wang
  et~al.}{2016a}]{Wang2016A}
Wang K.,  Guo P.,  Yin Q.,  Luo A.~L.,   Xin X.,  2016a, in International Joint
  Conference on Neural Networks. pp 3453--3460

\bibitem[\protect\citeauthoryear{Wang, Liu, Kumar  \& Chang}{Wang
  et~al.}{2016b}]{Wang2016Learning}
Wang J.,  Liu W.,  Kumar S.,   Chang S.,  2016b, Proceedings of the IEEE, 104,
  34

\bibitem[\protect\citeauthoryear{Wang, Guo, Luo, Xin  \& Duan}{Wang
  et~al.}{2017}]{Wang2017Deep}
Wang K.,  Guo P.,  Luo A.~L.,  Xin X.,   Duan F.,  2017, in IEEE International
  Conference on Systems, Man, and Cybernetics. pp 002687--002692

\bibitem[\protect\citeauthoryear{{Wei} et~al.,}{{Wei}
  et~al.}{2014}]{2014AJ....147..101W}
{Wei} P.,  et~al., 2014, \mn@doi [\aj] {10.1088/0004-6256/147/5/101}, \href
  {http://adsabs.harvard.edu/abs/2014AJ....147..101W} {147, 101}

\bibitem[\protect\citeauthoryear{Weiss, Torralba  \& Fergus}{Weiss
  et~al.}{2008}]{Weiss2008Spectral}
Weiss Y.,  Torralba A.,   Fergus R.,  2008, in International Conference on
  Neural Information Processing Systems. pp 1753--1760

\bibitem[\protect\citeauthoryear{{West} et~al.,}{{West}
  et~al.}{2011}]{2011ASPC..448.1407W}
{West} A.~A.,  et~al., 2011, in {Johns-Krull} C.,  {Browning} M.~K.,   {West}
  A.~A.,  eds,  Astronomical Society of the Pacific Conference Series Vol. 448,
  16th Cambridge Workshop on Cool Stars, Stellar Systems, and the Sun. p.~1407
  (\mn@eprint {arXiv} {1012.3766})

\bibitem[\protect\citeauthoryear{{Yang} et~al.,}{{Yang}
  et~al.}{2017}]{2017ApJ...849...36Y}
{Yang} H.,  et~al., 2017, \mn@doi [\apj] {10.3847/1538-4357/aa8ea2}, \href
  {http://adsabs.harvard.edu/abs/2017ApJ...849...36Y} {849, 36}

\bibitem[\protect\citeauthoryear{{Yanny} et~al.,}{{Yanny}
  et~al.}{2009}]{2009AJ....137.4377Y}
{Yanny} B.,  et~al., 2009, \mn@doi [\aj] {10.1088/0004-6256/137/5/4377}, \href
  {http://adsabs.harvard.edu/abs/2009AJ....137.4377Y} {137, 4377}

\bibitem[\protect\citeauthoryear{{Yi} et~al.,}{{Yi}
  et~al.}{2014}]{2014AJ....147...33Y}
{Yi} Z.,  et~al., 2014, \mn@doi [\aj] {10.1088/0004-6256/147/2/33}, \href
  {http://adsabs.harvard.edu/abs/2014AJ....147...33Y} {147, 33}

\bibitem[\protect\citeauthoryear{Zhang}{Zhang}{1999}]{Zhang1999An}
Zhang T.,  1999, Acm Sigmod Record, 25, 103

\bibitem[\protect\citeauthoryear{{Zhong} et~al.,}{{Zhong}
  et~al.}{2015}]{2015RAA....15.1154Z}
{Zhong} J.,  et~al., 2015, \mn@doi [Research in Astronomy and Astrophysics]
  {10.1088/1674-4527/15/8/005}, \href
  {http://adsabs.harvard.edu/abs/2015RAA....15.1154Z} {15, 1154}

\makeatother
\end{thebibliography}

\begin{landscape}
 \begin{table}
  \caption{Several examples of the online catalog\tnote{$^{a}$}$^{a}$\tnote{$^{b}$}$^{b}$.}
  \label{tab:table2}
 \begin{threeparttable}
  \begin{tabular}{cccccccccccccccc}
    \hline
    designation & obsDate & mjd & planID & spID & fiberID & ra & dec & snrr & subClass & rv & ewHa & ewHaErr & TiO5 & CaH2 & CaH3\\
    \hline
J005327.82+391733.4 & 2011-10-24 & 55858 & M5901 & 7 & 186 & 13.365946 & 39.292626 & 4.03 & M0 & -48.56 & 1.01 & 0.217 & 0.88 & 0.87 & 1.05\\
J005251.32+384930.7 & 2011-10-24 & 55858 & M5901 & 7 & 216 & 13.213834 & 38.825212 & 3.48 & M0 & -33.14 & 0.38 & 0.227 & 0.89 & 0.77 & 0.98 \\
J005009.21+382629.5 & 2011-10-24 & 55858 & M5901 & 7 & 244 & 12.538397 & 38.44155 & 3.15 & M0 & -51.63 & 1.44 & 0.333 & 0.76 & 0.74 & 0.91 \\
J005223.75+405459.0 & 2011-10-24 & 55858 & M5901 & 9 & 132 & 13.09896 & 40.916401 & 2.73 & M2 & 441.6 & -1.68 & 0.198 & 0.9 & 0.78 & 0.83 \\
J003718.31+394912.3 & 2011-10-24 & 55858 & M5901 & 10 & 172 & 9.3263115 & 39.820105 & 2.08 & M0 & 35.45 & 0.49 & 0.191 & 0.81 & 0.53 & 0.76 \\
    \hline
  \end{tabular}
 \end{threeparttable}
 \end{table}

 \begin{table}
 \begin{threeparttable}
  \begin{tabular}{ccccccccccccccccc}
    \hline
TiO1 & TiO2 & TiO3 & TiO4 & CaH1 & CaOH & TiO5Err & CaH2Err & CaH3Err & TiO1Err & TiO2Err & TiO3Err & TiO4Err & CaH1Err & CaOHErr & na & zeta \\
    \hline
0.87 & 0.9 & 0.99 & 0.89 & 0.89 & 0.79 & 0.044 & 0.037 & 0.044 & 0.051 & 0.06 & 0.056 & 0.051 & 0.03 & 0.044 & 1.03 & 2.56 \\
1 & 0.88 & 0.92 & 0.88 & 1 & 0.76 & 0.051 & 0.038 & 0.047 & 0.067 & 0.069 & 0.059 & 0.057 & 0.038 & 0.047 & 0.98 & 1.12 \\
1 & 1.03 & 0.95 & 0.79 & 0.91 & 0.78 & 0.056 & 0.048 & 0.058 & 0.086 & 0.106 & 0.082 & 0.067 & 0.047 & 0.058 & 1.14 & 1.71 \\
0.97 & 0.85 & 1.07 & 1.14 & 1.16 & 1.35 & 0.041 & 0.03 & 0.032 & 0.054 & 0.052 & 0.054 & 0.057 & 0.042 & 0.032 & 0.96 & 0.58 \\
0.72 & 0.77 & 0.9 & 0.94 & 0.82 & 0.52 & 0.035 & 0.02 & 0.028 & 0.038 & 0.044 & 0.043 & 0.046 & 0.031 & 0.028 & 0.92 & 0.45 \\
    \hline
  \end{tabular}
 \end{threeparttable}
 \end{table}

 \begin{table}
 \begin{threeparttable}
  \begin{tabular}{cccccccccccccccc}
    \hline
zetaerr & giant & parallax & parallax\_error & pmra & pmra\_error & pmdec & pmdec\_error & distance & U & V & W & h\_disc & l & b
 \\
    \hline
0.93 & 0 & -9999 & -9999 & -9999 & -9999 & -9999 & -9999 & -10 & -9999 & -9999 & -9999 & -12.29 & 132.040235 & 23.486443\\
0.62 & 0 & 0.6844 & 0.2827 & 5.783 & 0.553 & 2.885 & 0.319 & 1461.13 & 21.33 & -40.31 & 31.97 & -660.1 & 109.020028 & -59.220041\\
 0.658 & 0 & 2.0459 & 0.4649 & -0.554 & 0.775 & -1.613 & 0.712 & 488.78 & -27.3 & -40.17 & 17.96 & 480.95 & 93.931123 & -61.426298\\
 0.268 & 0 & -9999 & -9999 & -9999 & -9999 & -9999 & -9999 & -9999 & -9999 & -9999 & -9999 & -9999 & -9999 & -9999\\
 0.089 & 0 & 1.2663 & 0.1689 & 9.599 & 0.241 & -8.694 & 0.299 & 789.7 & 38.95 & -0.39 & -45.72 & -644.74 & 112.827489 & -71.309437\\
    \hline
  \end{tabular}
\begin{tablenotes}
\item[$^{a}$] The fields of the catalog are too many to be shown in one line, so they are split into several sub-tables. 
\item[$^{b}$] This table is just an example of the first four lines chosen from the complete catalog, more records can be found in the online catalog. 
\end{tablenotes}
 \end{threeparttable}
 \end{table}
\end{landscape}

\bsp	
\label{lastpage}
\end{document}